\documentstyle[12pt,epsf]{article}  
\def\fo{\hbox{{1}\kern-.25em\hbox{l}}}

\def\beq{\begin{equation}}
\def\eeq{\end{equation}}
\def\eq{\end{equation}}
\def\to{\rightarrow}

\def\bsg{\ifmmode B\to X_s\gamma\else $B\to X_s\gamma$\fi}
\def\bsll{\ifmmode B\to X_s\ell^+\ell^-\else $B\to X_s\ell^+\ell^-$\fi}
\def\bstt{\ifmmode B\to X_s\tau^+\tau^-\else $B\to X_s\tau^+\tau^-$\fi}
\def\shat{\ifmmode \hat{s}\else $\hat{s}$\fi}

\def\Emisst{\not  \! \! E_T}
\def\Emiss{\not  \! \! E}
\def\jet{{\rm jet}}

\newcommand{\newc}{\newcommand}

\newc{\lcal}{\int {\cal L}dt}
 
\newc{\LSP}{{\chi^0_1}}
\newc{\stauR}{{\tilde \tau_R}}
\newc{\stau}{{\tilde \tau_1}}
\newc{\mstop}{m_{\tilde{t}}}
\newc{\mHpm}{m_{H^\pm}}
\newc{\gsim}{\lower.7ex\hbox{$\;\stackrel{\textstyle>}{\sim}\;$}}
\newc{\lsim}{\lower.7ex\hbox{$\;\stackrel{\textstyle<}{\sim}\;$}}
\newc{\ie}{{\it i.e.}}          
\newc{\etal}{{\it et al.}}
\newc{\eg}{{\it e.g.}}          
\newc{\kev}{\hbox{\rm\,keV}}            
\newc{\mev}{\hbox{\rm\,MeV}}            
\newc{\gev}{\hbox{\rm\,GeV}}            
\newc{\tev}{\hbox{\rm\,TeV}}
\newc{\xpb}{\hbox{\rm\, pb}}
\newc{\xfb}{\hbox{\rm\, fb}}

\newc{\mtop}{m_t}
\newc{\mbot}{m_b}
\newc{\mz}{m_Z}
\newc{\mw}{M_W}
\newc{\alphasmz}{\alpha_s(m_Z^2)}
\newc{\swsq}{\sin^2\theta_W}
\newc{\tw}{\tan\theta_W}
\newc{\cw}{\cos\theta_W}
\newc{\sw}{\sin\theta_W}
\newc{\BR}{\hbox{\rm BR}}
\newc{\zbb}{Z\to b\bar}
\newc{\Gb}{\Gamma (Z\to b\bar b)}
\newc{\Gh}{\Gamma (Z\to \hbox{\rm hadrons})}
\newc{\rbsm}{R_b^\hbox{\rm sm}}
\newc{\rbsusy}{R_b^\hbox{\rm susy}}
\newc{\drb}{\delta R_b}

\newc{\sgn}{\mbox{sgn}}
\newc{\tbeta}{\tan\beta}
\newc{\uL}{{\tilde u_L}}
\newc{\uR}{{\tilde u_R}}
\newc{\cL}{{\tilde c_L}}
\newc{\cR}{{\tilde c_R}}
\newc{\tL}{{\tilde t_L}}
\newc{\tR}{{\tilde t_R}}
\newc{\dL}{{\tilde d_L}}
\newc{\dR}{{\tilde d_R}}
\newc{\sL}{{\tilde s_L}}
\newc{\sR}{{\tilde s_R}}
\newc{\bL}{{\tilde b_L}}
\newc{\bR}{{\tilde b_R}}
\newc{\eL}{{\tilde e_L}}
\newc{\eR}{{\tilde e_R}}
\newc{\mhp}{m_{H^\pm}}
\newc{\mhalf}{m_{1/2}}
\newc{\emt}{{e/\mu /\tau}}

\newc{\lR}{\tilde{l}_R}
\newc{\lL}{\tilde{l}_L}
\newc{\nL}{\tilde{\nu}_L}
\newc{\na}{\chi^0_1}
\newc{\nb}{\chi^0_2}
\newc{\nc}{\chi^0_3}
\newc{\nd}{\chi^0_4}
\newc{\ca}{\chi^{\pm}_1}
\newc{\cb}{\chi^{\pm}_2}
\newc{\camp}{\chi^\mp_1}
\newc{\cbmp}{\chi^\mp_1}
\newc{\capos}{\chi^{+}_1}
\newc{\caneg}{\chi^{-}_1}
\newc{\phit}{\phi_t}
\newc{\phib}{\phi_b}
\newc{\phiew}{\phi_{ew}}
\newc{\htz}{h^0_t}
\newc{\hbz}{h^0_b}
\newc{\hewz}{h^0_{ew}}
\newc{\hsmz}{h^0_{sm}}
\newc{\huz}{h^0_u}
\newc{\hsusyz}{h^0_{susy}}

\def\mp{M_P}
\def\ms{{\bar M}_P}
\def\mdr{{\bar M}_D}
\def\md{M_D}
\def\hn{{h^{(n)}}}
\def\sn{{S^{(n)}}}
\def\vn{{V^{(n)}}}
\def\gn{{G^{(n)}}}
\def\qn{{Q^{(n)}}}
\def\pn{{P^{(n)}}}

\def\hhn{{H^{(\vec n)}}}

\def\snv{{S^{(\vec n)}}}
\def\vnv{{V^{(\vec n)}}}
\def\gnv{{G^{(\vec n)}}}

\def\hhnv{{H^{(\vec n)}}}
\def\n{\hat{n}}
\def\kk{Kaluza-Klein }
\def\NPB#1#2#3{Nucl. Phys. B {\bf #1}, #3 (19#2)}
\def\PLB#1#2#3{Phys. Lett. B {\bf #1}, #3 (19#2)}

\def\PRD#1#2#3{Phys. Rev. D {\bf #1}, #3 (19#2)}

\def\PRT#1#2#3{Phys. Rep. {\bf#1} (19#2) #3}

\def\ZPC#1#2#3{Zeit. f\"ur Physik C {\bf #1}, #3 (19#2)}

\def\beq{\begin{equation}}
\def\eeq{\end{equation}}
\def\bea{\begin{eqnarray}}
\def\eea{\end{eqnarray}}
\def\slashchar#1{\setbox0=\hbox{$#1$}           
   \dimen0=\wd0                                 
   \setbox1=\hbox{/} \dimen1=\wd1               
   \ifdim\dimen0>\dimen1                        
      \rlap{\hbox to \dimen0{\hfil/\hfil}}      
      #1                                        
   \else                                        
      \rlap{\hbox to \dimen1{\hfil$#1$\hfil}}   
      /                                         
   \fi}                                         %
\catcode`@=11
\long\def\@caption#1[#2]#3{\par\addcontentsline{\csname
  ext@#1\endcsname}{#1}{\protect\numberline{\csname
  the#1\endcsname}{\ignorespaces #2}}\begingroup
    \small
    \@parboxrestore
    \@makecaption{\csname fnum@#1\endcsname}{\ignorespaces #3}\par
  \endgroup}
\catcode`@=12

\def\jfig#1#2#3{
 \begin{figure}
 \centering
 \epsfysize=4.6in
 \epsffile{#2}
 \caption{#3}
 \label{#1}
 \end{figure}}

\begin{document}

\baselineskip=18pt

\begin{titlepage}
\begin{flushright}
CERN-TH/98-354\\
hep-ph/9811291 
\end{flushright}

\begin{center}
\vspace{1cm}

{\Large \bf 

Quantum Gravity and Extra Dimensions \\
at High-Energy Colliders}

\vspace{0.5cm}

{\bf Gian F. Giudice, Riccardo Rattazzi, and James D. Wells }

\vspace{.8cm}

{\it Theory Division, CERN\\
     CH-1211 Geneva 23, Switzerland}

\end{center}
\vspace{1cm}

\begin{abstract}
\medskip

Recently it has been pointed out that the characteristic quantum-gravity
scale could be as low as the weak scale in theories with gravity
propagating in higher dimensions. The observed smallness of 
Newton's constant is a consequence of the large compactified volume
of the extra dimensions.
We investigate the consequences of this
supposition for high-energy collider experiments.  We do this by first
compactifying the higher dimensional theory and constructing a 
$3+1$-dimensional low-energy effective field theory of the graviton
Kaluza-Klein excitations and their interactions with
ordinary matter.  We then consider graviton production
processes, and select $\gamma+\Emiss$ and $\jet +\Emiss$ 
signatures for careful study.  We find that both a $1\tev$ $e^+e^-$
collider and the CERN LHC will be able to reliably and perturbatively
probe the fundamental gravity scale up to several TeV, with the
precise value depending on the number of extra dimensions.  
Similarly, searches at LEP2
and the Tevatron are able to probe this scale up to approximately $1\tev$.
We also
discuss virtual graviton exchange, which induces local 
dimension-eight operators associated with the square of the energy-momentum
tensor.  We estimate the size of
such operators and study their effects on $f\bar f\to
\gamma\gamma$ observables.  

\end{abstract}

\bigskip
\bigskip

\begin{flushleft}
CERN-TH/98-354 \\
November 1998
\end{flushleft}

\end{titlepage}

\tableofcontents

\section{Introduction}

The idea that quantum-gravity effects appear only at energy scales
near the Planck mass $\mp = 1.2\times 10^{19}$ GeV has been
shaken recently~\cite{dim,lyk,wit,hor,ant1}. Witten~\cite{wit} has
pointed out that the string scale $M_S$ can be substantially
lower than that predicted by
the tree-level relation~\cite{gins}
\beq
M_S =\mp \frac{\sqrt{k\alpha_{GUT}}}{2}.
\label{string}
\eeq
Here $\alpha_{GUT}$ is the unified gauge coupling constant and
$k$ is the Ka{\v c}-Moody 
level, typically of order unity. Equation~(\ref{string})
predicts a value of $M_S$ which is too large to be consistent with 
unification of the gauge couplings measured at low energies. 
Several solutions have been proposed to 
cure this phenomenologically unpleasant aspect
of string theory:  large threshold corrections,
high Ka{\v c}-Moody levels, or GUT models embedded in strings (for a review,
see ref.~\cite{die}). However, as pointed out in ref.~\cite{wit},
explicit calculations in the strong-coupling regime using string duality
show that eq.~(\ref{string}) receives large corrections, and $M_S$ can be 
lowered to values compatible with gauge coupling unification. Thus one
can claim to have achieved a complete unification of gauge and gravity
forces at a single energy scale,
although 
$M_S$ is now a model-dependent parameter, and therefore no new low-energy
predictions can be derived. 

The result that the string and the Planck scale are not necessarily
tied together by eq.~(\ref{string}) has been pushed to its extreme consequences
by Lykken~\cite{lyk}. Indeed, even in the absence of realistic models, one
can speculate on the radical possibility that the string scale is as low as the
Fermi scale. If this is the case, one loses the original motivation for
space-time supersymmetry, based on the hierarchy problem. Supersymmetry
may still be desirable as a necessary ingredient of string theory, but it could
be broken at the string level and not be present in the effective low-energy
field theory. For collider experiments this implies that the superparticle
masses should follow a pattern of string Regge recurrences.

Arkani-Hamed, Dimopoulos, and Dvali~\cite{dim} have made the very interesting
proposal that, while Standard Model particles live in the usual
3+1-dimensional space, gravity can propagate in a higher-dimensional space.
The relative feebleness of gravity with respect to weak
forces is related to the 
size of the compactified extra dimensions which, in units
of TeV$^{-1}$, is very large. 
Newton's constant measured in the 3+1-dimensional space
can be expressed as~\cite{dim}
\beq
G_N^{-1} =8\pi R^\delta \md^{2+\delta},
\label{newt}
\eeq
where $\md \sim$ TeV is the fundamental mass scale, $\delta$ is the number
of extra dimensions, and $R$ is the radius of the compactified
space, here assumed to be a torus. 
The hierarchy problem is overcome, because there is a single fundamental 
mass scale
$\md$, to be identified with the weak scale.
Since gravity forces are not well probed at distances less than
a millimeter~\cite{gra}, eq.~(\ref{newt}) 
is consistent with present measurements
as long as $\delta$ is larger than or equal to 2. 
Deviations from the standard Newtonian law of gravitational
attraction are predicted
at distances smaller than $R\sim 10^{\frac{32}{\delta}-19}$
meters. In the case $\delta =2$, such effects will be probed at
future experiments, which are sensitive to gravitational forces
down to distances of tens of microns.

Although the idea that our 3+1-dimensional world could be a field-theoretical
topological defect of a higher-dimensional theory dates
back to ref.~\cite{sha}, it finds a natural setting
in the context of string theory. Indeed, Dirichlet branes are defects
intrinsic to string theory (for reviews, see ref.~\cite{bra}). 
Standard Model particles are naturally confined
to the lower-dimensional space, since they correspond to open 
strings with the endpoints attached to the brane. On the other hand,
gravitons correspond to closed strings which propagate in the whole
higher-dimensional space. This picture of ordinary particles confined
on the brane with gravity propagating in the bulk can be realized in
several string models~\cite{hor,dima,tye}.

Recently, there has been considerable activity~\cite{dima}--\cite{sun2}
in investigating the
proposal of ref.~\cite{dim} and related ones.
These scenarios are undoubtedly very intriguing, although 
there are still many theoretical open questions. 
First of all, 
the
solution to the hierarchy problem is not complete until we can predict
the radius $R$ of the compactified space. After all, the effect of
eq.~(\ref{newt}) is just to trade
the small ratio
$M_W/\mp$ with the large number $(RM_W)^{\delta /2}$. Interesting 
attempts to address this problem have been presented in 
refs.~\cite{sun1,dimr}. The second problem we want to mention is
related to the cosmology of this scenario. During the early phase of the
universe, energy can be emitted from the brane into the bulk in the form
of gravitons. The gravitons propagate in the extra dimensions and can
decay into ordinary particles only by interacting with the brane, and
therefore with a rate suppressed by $1/\mp^2$. Their contribution to
the present energy density exceeds the critical value unless~\cite{dim}
\beq
T_\star < \frac{\md}{\rm TeV}~ 10^{\frac{6\delta -15}{\delta +2}}{\rm MeV}.
\label{tstar}
\eeq
Here $T_\star$ is the maximum temperature 
to which we can simply extrapolate the thermal history of
the Universe, assuming it is in a 
stage with completely stabilized $R$ and with vanishing energy density
in the compactified space. 
As a possible example of its origin, 
$T_\star$ could correspond to the reheating temperature
after an inflationary epoch.
The bound in
eq.~(\ref{tstar}) is very constraining. In particular,
for $\delta =2$, only values of $\md$ larger than about 6 TeV can lead
to $T_\star > 1$ MeV and allow for standard nucleosynthesis. Moreover,
even for larger values of $\delta$, eq.~(\ref{tstar}) is
problematic for any mechanism of baryogenesis~\cite{dav}. 

The possibility that ordinary matter lives
in a higher-dimensional space 
with TeV$^{-1}$ size
has also been considered in the literature~\cite{ant1,ant2}.
Low-energy supersymmetry breaking within perturbative string theory
then can be related to the radius of the compactified space.
Such a point of view is not inconsistent with the previous picture. Our
world would be confined to a $d$-dimensional space with $d-4$ coordinates
of size $r\sim$ TeV$^{-1}$, which is embedded in a $D$-dimensional
space with extra coordinates of size $R$ (much larger than $r$) where
only gravity is free to propagate. Recently, it has been argued that
the \kk states of the Standard Model particles could be used to lower
the GUT scale~\cite{ghe}.

In summary, it does not appear implausible that quantum-gravity effects
can start revealing themselves at energies much lower than $\mp$, 
possibly as low as the weak scale. This has the exciting implication 
that future high-energy collider experiments can directly probe the
physics of quantum gravity. Above the TeV energy scale, completely new
phenomena could emerge as  resonant production of the Regge recurrences
of string theory or excitations of \kk modes of ordinary 
particles~\cite{qui}. 

There is little doubt that, if these scenarios have some truth, the
collider phenomenology above the TeV scale would be quite distinct from
Standard Model expectations. However, because of our basic ignorance about
the underlying quantum-gravity theory, it is less clear what the
distinguishing experimental signatures would be,
and whether any of these signatures
would depend only on the conceptual theoretical
hypothesis and not on the specific model
realization. These are the questions we want to address in this paper.

We consider the 
scenario of ref.~\cite{dim}, in which gravity 
can propagate into extra dimensions
and define an effective theory valid below the fundamental scale $\md$.
As usual, the contribution of the unknown ultraviolet physics is reabsorbed
in unknown coefficients of the effective theory. However,
the advantage of the
effective-theory approach is that it allows one to separate ultraviolet from
infrared contributions and therefore to derive some model-independent
results, which are affected only by the infrared behaviour. In particular,
we will argue that rates for  graviton production 
are model independent, as long as the typical energy is less
than $\md$.

Let us consider the form of the effective theory below $\md$. The dynamical
degrees of freedom are described by the Standard Model particles, by the
graviton, and by possible other light fields related to the brane 
dynamics~\cite{sun2}.
In particular, the $Y$ modes describing the deformations of the brane 
in the $D$-dimensional space could be much lighter than $\md$ or even
massless, if the brane breaks 
translation invariance in the transverse
directions only spontaneously. 
The modes $Y$ are coupled to ordinary matter only in pairs and
therefore their production cross-section is subleading with respect to
the case of gravitons, which are singly produced. Thus we will disregard
these fields. Indeed, for simplicity, we can just assume that the
brane is rigid with its position fixed, and the $Y$ fields 
have masses of order $\md$ or larger. For instance,
this case arises when the Standard Model 
degrees of freedom live at an  orbifold
singularity.


The graviton corresponds to the excitations of the $D$-dimensional metric.
In terms of 4-dimensional indices, the metric tensor contains spin-2, spin-1,
and spin-0 particles. 
Moreover, since these fields depend on $D$-dimensional coordinates, they 
can be expressed as a tower of \kk modes. The mass of each \kk
mode corresponds to the modulus of its momentum in the direction
transverse to the brane. The picture of a massless graviton propagating
in $D$ dimensions and the picture of massive \kk gravitons propagating in
4 dimensions are equivalent,  and we will often use both descriptions
in our discussion.

In the low-energy limit, the coupling of the Kaluza-Klein gravitons with the
particles on the brane is determined by general covariance both in
the full $D$-dimensional theory and in the 4-dimensional brane description.
At lowest order this means that the metric on the brane
is simply the bulk metric projected to the 4 dimensions parallel
to the brane. An important consequence is then the universal nature
of the coupling of the \kk modes.


At lowest order in $1/\mp$ the above remark is 
sufficient to predict the emission rates
of real gravitons in the effective theory. In this way we can calculate
the experimental signal expected at high-energy colliders and 
can compare it with the Standard Model background. This provides a 
rather model-independent test of the idea that gravity can propagate in
extra dimensions. Effectively here we are studying quantum gravity in
its weak-coupling regime, where we can make definite predictions, 
and we are determining
its behaviour before the onset of the fundamental underlying theory.

It may seem hopeless to observe processes with real graviton emission,
since the relevant interaction is suppressed by inverse powers of
$\mp$. However, the large phase space of the \kk modes, corresponding
to the large volume of the compactified space exactly cancels the
dependence on $\mp$ and gives an effective interaction suppressed
only by inverse powers of $\md$ (see sect.~\ref{seccros}). In the
$D$-dimensional language this is evident, since the graviton interactions
are determined by the only available scale $\md$. Therefore, for 
collider applications, we can work in the limit of infinite compactified
space ($R\to \infty$) in which the graviton \kk excitations have a
continuous spectrum, and the usual four-dimensional gravitational effects
are shut off.

An important remark is that we have to be aware that other effects
inherent to the fundamental theory, and therefore not computable with
an effective Lagrangian approach, can give various
experimental signals.  These effects could be more easily detectable
than the effects we are studying. Therefore the discovery
modes could be different than what is discussed here. However, these
will be model-dependent effects, and little can be said about them with
sufficient generality at
present. The specific experimental processes discussed here can provide
a handle to disentangle unexpected signals and test a precise hypothesis.
Moreover, in case no deviation from the Standard Model
is observed, they define in a quantitative
way the strategy to obtain lower bounds on the new physics energy scale.

Another remark concerns the value of the ultraviolet validity cutoff
of the effective theory. In practice, this is an important issue,
because this cutoff determines the maximum energy to which we can
extrapolate our predictions and, analogously, the minimum value of
$\md$ which can be studied reliably at a collider experiment. This
energy cutoff is expected to be of order $\md$. As we will show in
sect.~4 and 5, naive dimensional analysis and unitarity arguments
suggest
that the effective theory could be valid up to
energies quite larger than $\md$. In reality it is more
reasonable to believe that
the fundamental theory that regularizes quantum gravity sets in well before
the latter becomes strongly interacting. In the context of string
theory, the belief is that the string scale is smaller than $\md$, and
therefore our effective theory has a more limited applicability energy range,
although large enough to be used for collider predictions, as we will
illustrate in the sect.~\ref{seccol}.
It is also possible that the fundamental theory of gravity introduces
new phenomena at scales equal to or less than $\md$ but does not
significantly corrupt the graviton-emission signals up to larger energy
scales.

The remainder of the paper is organized as follows. In sect.~2 we introduce
the graviton \kk modes and identify the physical degrees of freedom.
In sect.~3 we derive the graviton Feynman rules necessary for our
calculation. The cross-sections for graviton production and for
processes mediated by virtual-graviton exchange are computed in sects.~4
and 5. The analysis of the observability of graviton signals at
future high-energy colliders is contained in sects.~6 and 7. 

\section{The \kk Excitations of the Graviton}
\label{seckk}

In this section we study the equations that describe 
the \kk excitations of the graviton and identify the physical degrees
of freedom in the effective theory. At low energy and small curvature
the equations of motion of the effective theory reduce to the
Einstein equation in $D=4+\delta$ dimensions
\beq
{\cal G}_{AB}\equiv
{\cal R}_{AB}-\frac{1}{2} g_{AB}{\cal R}
=-\frac{T_{AB}}{\mdr^{2+\delta}}~~~~~~~A,B=1,\dots ,D,
\label{einst}
\eeq
where $\mdr$ 
is the reduced Planck mass of the $D$-dimensional theory. 

In general, the presence of the brane on which we live will
create a non-trivial D-dimensional metric background. However, it is 
quite reasonable to expect that the brane surface tension $f^4$ 
does not exceed the fundamental scale
$M_D^4$. Therefore, when the distance from the brane gets much
bigger that $ 1/M_D$, the metric $g_{AB}$ will be essentially
flat. Correspondingly, if we study the emission of ``soft'' gravitons,
with a momentum transverse to the brane $q_T\ll M_D$, we are  only 
concerned with distances at which the metric is essentially 
flat\footnote{We thank Raman Sundrum for important comments on this 
issue, see also ref.~\cite{sun1}.}.
In view of the above remark, 
we expand the metric $g_{AB}$ around its Minkowski value $\eta_{AB}$
\beq
g_{AB}=\eta_{AB}+2\mdr^{-1-\delta /2} h_{AB}.
\label{metrexp}
\eeq
Keeping only the first power of $h$, eq.~(\ref{einst}) becomes
\bea
\mdr^{1+\delta /2} {\cal G}_{AB}&=& 
\Box h_{AB}-\partial_A \partial^Ch_{CB}
-\partial_B \partial^Ch_{CA}
+\partial_A \partial_B h^C_{C}\nonumber \\
&-&\eta_{AB}\Box h^C_{C} + \eta_{AB}
\partial^C \partial^Dh_{CD} 
=-\mdr^{-1-\delta /2}
T_{AB} ,
\label{weak}
\eea
where indices are raised or lowered using the flat-space metric and
summation over repeated indices is understood. 

A point in the
$D$-dimensional space is described by the set of coordinates 
$z=(z_1,\dots ,z_D)$.
Let us explicitly separate the ordinary four-dimensional coordinates
{}from the extra dimensional ones, 
\beq
z=(x,y)~~~~~~x=(x_0,\vec{x}),~y=(y_1,\dots ,y_\delta ),~~~~\delta=D-4 .
\eeq
We now demand periodicity of the fields under the following
translation in the compactified
space
\beq
y_j\to y_j +2\pi R ~~~~j=1,\dots ,\delta ,
\label{period}
\eeq
where $R$ is the compactification radius. For simplicity we assume that
the compactified space is a torus, but our considerations are valid
for different
compact spaces. 
Equation~(\ref{period}) implies
\beq
h_{AB}(z)=\sum_{n_1=-\infty}^{+\infty} \dots \sum_{n_\delta=-\infty}^{+\infty}
\frac{\hn_{AB}(x)}{\sqrt{V_\delta}}e^{i\frac{n^jy_j}{R}},
\eeq
where $n=(n_1,\dots ,n_\delta)$ and $V_\delta$ is the volume
of the compactified space,
\beq
V_\delta =(2\pi R)^\delta .
\label{vol}
\eeq
The tensor $h(z)$ has been split into an
infinite sum of \kk modes $\hn (x)$ which live in the 
four-dimensional space.

We assume that ordinary matter is confined on the brane, and
therefore, in the limit of weak gravitational field, the energy-momentum
tensor becomes
\beq
T_{AB}(z)=\eta_A^\mu \eta_B^\nu T_{\mu \nu}(x) \delta(y) ~~~~\mu,\nu=0,\dots ,3.
\label{tensor}
\eeq
The singularity of the $\delta$ function in 
eq.~(\ref{tensor}) will presumably
be smoothed by effects of the finite brane-size in
the transverse direction. 
However, these effects are important only in the short-distance regime.
For our purposes, 
eq.~(\ref{tensor}) means that
the \kk modes of the energy-momentum tensor are independent
of $n$, in the low-energy region, where $n$ is smaller than $\mdr R$. This
is a crucial ingredient of our analysis, since it entails a universal
coupling of all relevant \kk gravitons to ordinary matter, thus
allowing us to make definite predictions on their production cross-sections.

Notice that eq.~(\ref{tensor}) can more formally be obtained via the
induced metric $\hat g_{\mu\nu}$ on the brane
\beq
\hat g_{\mu\nu}(x)=g_{AB}[Y(x)]\partial_\mu Y^A(x)\partial_\nu Y^B(x)
\eeq
where $Y^A$ are the background fields describing the position of
the brane. The metric $\hat g$ measures distances on the brane and
should be used to write a covariant action.
In the static gauge ($Y^\mu=x^\mu$ for $\mu=0,\dots,3$
and $Y_i=0$ for $i=4,\dots,D-1$) one simply has $\hat g_{\mu\nu}=g_{\mu\nu}$,
and the above coupling to bulk $(\mu,\nu)$ gravitons follows.

We should recall that in general there could also be couplings
to $(i,j)$ gravitons (polarized orthogonal to the brane). 
In particular, the fields $Y^i$, if dynamical, couple to gravitons
via $h_{ij}\partial_\mu Y^i\partial_\nu Y^j \eta^{\mu\nu}$. Moreover,
in the case of D-branes there are also fermions (the partners of
$Y^i$) that couple to the perpendicularly polarized gravitons $h_{ij}$,
while gauge bosons obviously cannot. The coupling to $h_{ij}$
is evidently model dependent, and it does not seem inconsistent to assume 
that it vanishes for the Standard Model 
degrees of freedom. This is the assumption underlying
eq.~(\ref{tensor}) and the rest of our analysis.

In the context of D-branes, there exist interesting papers discussing
the emission and absorption of gravitons by the brane \cite{kle1,kle2}.
In this case, using string theory, one can even perform the calculation in the high-energy
region, where $\sqrt {s}$ is larger than the string scale.  It is worth pointing out that 
the low-energy (soft gravitons) limit of that calculation agrees 
with the result obtained by an effective theory, 
 where the bulk gravitons propagate in a flat background 
and couple to the fields on the brane via the induced metric~\cite{kle1}.

Let us now go back to the equations of motion.
After multiplying 
both sides of the \kk expansion of eq.~(\ref{weak}) 
by $e^{-i\frac{n'}{R}\cdot y}$
and integrating over the extra-dimensional
coordinates, we obtain the following set of equations:
\bea
{{\cal G}^{(n)}}_{\mu \nu}(x)&\equiv &
 ( \Box +\n^2) \hn_{\mu \nu} - 
\left[ \partial_\mu \partial_\lambda \hn_\nu^\lambda 
+i\n_j \partial_\mu \hn_\nu^j
+ (\mu \leftrightarrow \nu )\right]
+\nonumber \\
&&\left[ \partial_\mu \partial_\nu -\eta_{\mu \nu} (\Box +\n^2)\right]
\left[ \hn_\lambda^\lambda + \hn_j^j\right] +\nonumber \\
&& \eta_{\mu \nu}\left[
\partial^\lambda \partial^\sigma \hn_{\lambda \sigma} +2i\n_j \partial^\lambda
\hn_\lambda^j -\n^j\n^k \hn_{jk} \right]
=-\frac{T_{\mu \nu}}{\ms} ,
\label{ein1}
\eea
\bea
 {{\cal G}^{(n)}}_{\mu j}(x)&\equiv &
( \Box +\n^2) \hn_{\mu j} - 
\partial_\mu \partial_\nu \hn_j^\nu
-i \n_k \partial_\mu \hn_j^k
-i \n_j \partial_\nu \hn_\mu^\nu
\nonumber \\ &&
+\n_j\n_k \hn_\mu^k 
+i\n_j\partial_\mu \left[ \hn_\nu^\nu + \hn_k^k \right]
=0 ,
\label{ein2} 
\eea
\bea
{{\cal G}^{(n)}}_{jk}(x)&\equiv  &
( \Box +\n^2) \hn_{ jk} - 
\left[ i\n_j \partial_\mu \hn_k^\mu
- \n_j \n_\ell \hn_k^\ell    + (j \leftrightarrow k)\right]\nonumber \\
&&-\left[ \n_j \n_k +\eta_{jk} (\Box +\n^2)\right]
\left[ \hn_\mu^\mu + \hn_\ell^\ell\right] +\nonumber \\
&&\eta_{jk}\left[
\partial^\mu \partial^\nu \hn_{\mu \nu} +2i\n_\ell
\partial^\mu \hn_\mu^\ell -\n^\ell \n^m \hn_{\ell m} \right] =0.
\label{ein3}
\eea
Here 
the D'Alambertian operator
acts on the four-dimensional space $\Box =\partial^\mu
\partial_\mu$, and we have defined\footnote{In our 
conventions, the flat metric is $\eta_{\mu \nu}
={\rm diag}(+,-,-,-)$ and $\eta_{jk}=-\delta_{jk}$.}
 $\n \equiv n/R$, $\n^2\equiv -\n^j\n_j=\sum_{j=1}^\delta 
|\n_j|^2$.
We can now interpret the quantity
\beq
\ms \equiv \sqrt{V_\delta} \mdr^{1+\delta /2} = (2\pi R)^{\delta /2}
\mdr^{1+\delta /2}\equiv R^{\delta /2}\md^{1+\delta /2}
\label{planc}
\eeq
as the
ordinary reduced Planck mass,
$\ms=\mp /\sqrt{8\pi}=2.4\times 10^{18}$ GeV. For future convenience
we have also defined the $D$-dimensional Planck mass, related to the
reduced Planck mass, by the equation $\md = (2\pi)^{\delta /(2+\delta)}\mdr$.
With eq.~(\ref{planc}) we have rederived, using the point of view of 
general relativity, the relation~(\ref{newt}) between $\mp$ and $\md$
previously obtained in ref.~\cite{dim}.

In order to solve the system of coupled differential equations 
it is convenient to rewrite them in terms
of the following new dynamical variables:
\bea
\gn_{\mu \nu} &\equiv & \hn_{\mu \nu} 
+\frac{\kappa}{3}\left( \eta_{\mu \nu} +\frac{\partial_\mu \partial_\nu}{\n^2}
\right) \hhn
- \partial_\mu \partial_\nu
\pn + \partial_\mu \qn_\nu + \partial_\nu \qn_\mu 
\label{defg}
\\
\vn_{\mu j} &\equiv & \frac{1}{\sqrt{2}}\left[
i \hn_{\mu j} -\partial_\mu \pn_j -\n_j \qn_\mu \right] \\
\sn_{jk} &\equiv & \hn_{jk} 
-\frac{\kappa}{\delta -1}\left( \eta_{jk}+\frac{\n_j\n_k}{\n^2}\right) \hhn
+\n_j \pn_k +\n_k \pn_j -\n_j \n_k \pn \\
\hhn &\equiv & \frac{1}{\kappa} \left[ \hn_j^j +\n^2 \pn \right] \\
\qn_\mu &\equiv & -i \frac{\n_j}{\n^2}\hn_\mu^j \\
\pn_j &\equiv &\frac{\n_k}{\n^2}\hn_j^k+\n_j \pn \\
\pn &\equiv & \frac{\n^j \n^k}{\n^4}\hn_{jk} .
\label{defp}
\eea
The degrees of freedom contained in $\gn_{\mu \nu},\vn_{\mu j},
\sn_{jk},\hhn ,
\qn_\mu,\pn_j,\pn$ correctly match the number of degrees of freedom 
in $h_{AB}$ without giving a redundant description, because of the identities 
$\n^j \vn_{\mu j}=0$, $\n^j \sn_{jk}=0$, $\sn_j^j=0$, $\n^j \pn_j =0$.
For convenience, we choose
\beq
\kappa =\sqrt{\frac{3(\delta -1)}{\delta +2}},
\label{cappa}
\eeq
in order to have a canonical normalization of the field $\hhn$. Notice that,
for $\delta =1$,
our parametrization is singular, since the fields $\pn_j$, $\pn$, and
$\hhn$ are no longer independent. However, we will be interested only
in the case $\delta >1$. 

By contracting the free indices of eqs.~(\ref{ein1})--(\ref{ein3})
either with the flat metric
tensor
or with $\partial^\mu$ and $\n^j$ 
($j=1,\dots ,\delta$), we obtain 
three constraints\footnote{The constraints come from the equations
${{\cal G}^{(n)}}_{\mu}^\mu = -T^\mu_\mu /\ms$, 
$\partial^\mu {{\cal G}^{(n)}}_{\mu \nu}=0$,
and $\n^j {{\cal G}^{(n)}}_{jk}=0$.
Notice that the conditions
$\n^j {{\cal G}^{(n)}}_{\mu j}=0$ and $\partial^\mu 
{{\cal G}^{(n)}}_{\mu j}=0$ do not
provide further constraints because they identically follow from the
previous equations and  energy-momentum conservation in
$D$ dimensions. Finally the information of the equation ${{\cal G}^{(n)}}_j^j
=0$ is directly contained in the equations
of motion.}
on the \kk 
components of the tensor $h$ ($n\ne 0$):
\bea
\partial^\mu \gn_{\mu \nu}&=&\frac{\partial_\nu T^\mu_\mu}{3\n^2 \ms}
\label{h1} \\
\gn^\mu_\mu  &=& \frac{T^\mu_\mu}{3\n^2 \ms}
\label{h2} \\
\partial^\mu\vn_{\mu j} &=& 0. \label{cons}
\eea
Replacing in eqs.~(\ref{ein1})--(\ref{ein3}) the constraints given in 
eqs.~(\ref{h1})--(\ref{cons}), we can rewrite the equations
of motion for each of the \kk modes $n\ne 0$ as
\bea
&\left( \Box +\n^2\right) & \gn_{\mu \nu}=
\frac{1}{\ms} \left[ -T_{\mu \nu} +\left( \frac{\partial_\mu 
\partial_\nu}{\n^2} +\eta_{\mu\nu}\right) \frac{T^\lambda_\lambda}{3}
\right] \label{h3} \\
&\left( \Box +\n^2\right) & \vn_{\mu j}
=0 \label{hvec} \\
&\left( \Box +\n^2\right) & \sn_{jk} = 0 \label{h301} \\
&\left( \Box +\n^2\right) & \hhn = \frac{\kappa}{3\ms}T^\mu_\mu .
\label{h30}
\eea

These equations show that only $\gn_{\mu \nu},\vn_{\mu j},
\sn_{jk},\hhn$ describe propagating particles, while $\qn_\mu,\pn_j,\pn$ do not
appear in the equations of motions. This result can be understood by
studying the transformation properties of the different fields 
under coordinate reparametrization.
Let us consider 
a general coordinate transformation 
\beq
z_A \to z'_A=z_A+\epsilon_A(z),
\eeq
where $\partial_B \epsilon_A$ is at most of the same order of magnitude 
as $h_{AB}$. This transformation induces a variation of the metric such that
\beq
\delta_\epsilon h_{AB} =-\partial_A \epsilon_B -\partial_B \epsilon_A.
\label{gautra}
\eeq
Using a \kk mode expansion for the gauge parameter $\epsilon$
\beq
\epsilon_A (z) = 
\sum_{n_1=-\infty}^{+\infty} \dots \sum_{n_\delta=-\infty}^{+\infty}
{\epsilon^{(n)}}_A(x)e^{i\frac{n^jy_j}{R}},
\eeq
the diffeomorphism in eq.~(\ref{gautra}) induces the following
transformation laws for the fields 
defined in eqs.~(\ref{defg})--(\ref{defp}),
\beq
\delta_\epsilon \gn_{\mu \nu} =0,~~~~\delta_\epsilon \vn_{\mu j}=0
,~~~~\delta_\epsilon \sn_{jk} =0,~~~~\delta_\epsilon \hhn =0,
\eeq
\bea
\delta_\epsilon \qn_\mu &=&\frac{1}{2}\partial_\mu \delta_\epsilon 
\pn +{\epsilon^{(n)}}_\mu \\
\delta_\epsilon \pn_j &=& -\frac{1}{2} \n_j \delta \pn -i {\epsilon^{(n)}}_j \\
\delta_\epsilon \pn &=& 2i \frac{\n_j}{\n^2}{\epsilon^{(n)}}^j .
\eea
Thus, while $\gn_{\mu \nu},\vn_{\mu j},
\sn_{jk},\hhn$ 
are gauge-invariant fields, $\qn_\mu,\pn_j,\pn$ are gauge-dependent
and do not describe physical particles. In particular, 
they can all be simultaneously set to zero
at any four-dimensional space-time
point for any $n \ne 0$. We will refer to the gauge choice in which
$\qn_\mu =0$, $\pn_j=0$, $\pn=0$ as the unitary gauge.

We now want to identify the physical content of the gauge-invariant fields.
The
free propagation of $\gn_{\mu \nu}$
is given by eqs.~(\ref{h1}) and (\ref{h3}) in the limit $T_{\mu \nu}=0$,
\beq
\left( \Box +\n^2\right) \gn_{\mu \nu} =0
\label{hg1}
\eeq
\beq
\partial^\mu \gn_{\mu \nu}=0
\label{hg2}
\eeq
\beq
\gn^\mu_\mu =0.
\label{hg3}
\eeq
Equation ~(\ref{hg1}) describes the free propagation of the bosonic modes
$\gn_{\mu \nu}$ having squared masses equal to $\n^2$. For any given $n$,
the constraints in 
eqs.~(\ref{hg2})--(\ref{hg3}) eliminate 5 degrees of freedom out of
the 10 contained in the symmetric tensor $\gn_{\mu \nu}$. This leaves
5 propagating modes corresponding to the physical degrees of freedom 
of a massive spin-two particle, the $n$-th \kk excitation of the
graviton. The condition of the unitary gauge corresponds to 
eliminating the spin-zero ($\pn$) and spin-one ($\qn_\mu$)
particles eaten by the massless graviton to form a spin-two massive
multiplet. 

The fields $\vn_{\mu j}$ (satisfying $\n^j \vn_{\mu j}=0$)
describe $\delta -1$ spin-one particles which
form massive multiplets by absorbing the fields $\pn_j$. 
These vector particles satisfy
the Lorentz condition, see eq.~(\ref{cons}), and each contains three physical
degrees of freedom. They are not coupled to the energy-momentum tensor
in the weak-field limit, see eq.~(\ref{hvec}),
and therefore will play no role in our analysis
of  collider experiments. Next, for $\delta \geq 2$, 
there are $(\delta^2 -\delta -2)/2$ massive real
scalars described by the symmetric tensor $\sn_{jk}$, which satisfies
the relations $\n^j \sn_{jk}=0$, $\sn_j^j =0$.
These scalars are also not coupled to matter,
see eq.~(\ref{h301}). Finally, there is the scalar $\hhn$ which is
coupled only to the trace of the energy-momentum tensor, see eq.~(\ref{h30}).
If we impose the equations of motion, $T_\mu^\mu$ vanishes 
for conformally-invariant theories. 
Thus $\hhn$ does not participate in any tree-level process involving
massless matter field.
The scalar $\hhn$ can only couple to 
ordinary particles at tree level proportionally to their masses.
These couplings give effective interactions at best of order 
$M_Z^2/M_D^2$, which 
are negligible for our considerations. 
Indeed one may worry about the existence of a massless  mode associated 
with 
$\hhn$, which would lead to unwanted violations of the Equivalence Principle
(a difference in Newton's constant characterizing the coupling
to photons and to non-relativistic matter). The mode in question 
corresponds to a fluctuation, named {\it} radion in ref. \cite{dimr},
of the volume of the $\delta$ compactified dimensions. Thus, whatever 
mechanism stabilizes the radius of the extra dimensions, it will give a mass
to the radion. In ref. \cite{dimr} several mechanisms of 
radius stabilization were described and it was shown that it is
not difficult to give the radion a mass larger than $(1 ~{\rm mm})^{-1}
\sim 2\times 10^{-4}~{\rm eV}$. This is sufficient to satisfy the present experimental
bounds. Such a tiny mass for $\hhn$ affects our 
previous analysis only for the lowest \kk modes.
As a final remark, the radion corresponds to the zero mode of
$h^j_j$.
Since $n=0$,
this mode cannot be gauged away even in the special case of $\delta=1$,
in which the radion \kk excitations
are eaten by the
massive spin-2 graviton.

Adding the numbers of physical degrees of freedom for
$\gn_{\mu \nu}$, $\vn_{\mu j}$, 
$\sn_{jk}$, and $\hhn$
we obtain a total of $(4+\delta )(1+\delta )/2$. It is 
interesting to do the same 
counting from the point of view of the $D$-dimensional
theory. The symmetric tensor $h_{AB}$ contains $D(D+1)/2$ massless components.
In order to compute the physical degrees of freedom,
we first need
to fix the gauge, say by the harmonic condition $\partial_A h^A_B =\frac{1}{2}
\partial_B h^A_A$ (see {\it e.g.} ref.~\cite{wein}). 
Analogously to the case of QED, we then discover 
that, for a massless graviton, we still have residual freedom to make gauge
transformations with gauge parameters $\epsilon_A$ subject to
the condition  $\Box \epsilon_A =0$. In total we eliminate $2D$ components
{}from $h_{AB}$, leaving $D(D-3)/2$ physical 
states. For $D=4$ we find the 2 degrees
of freedom of a massless graviton, and for $D=4+\delta$ we recover the
same result previously obtained from a four-dimensional \kk point of view.

\section{Feynman Rules}
\label{secfey}

In this section we give the Feynman rules necessary to 
compute the rate for graviton-emission processes.
Some of the rules presented here can also be extracted from
ref.~\cite{feyn}. 

We start from the $D$-dimensional graviton 
Lagrangian
corresponding to the
Einstein equation (\ref{weak}), 
\beq
{\cal L}= -\frac{1}{2} h^{A B} \Box h_{AB}
+\frac{1}{2} h^A_A \Box h^B_B
-h^{AB}\partial_A \partial_B h^C_C 
+h^{AB}\partial_A \partial_C h^C_B -\frac{1}{\mdr^{1+\delta / 2}}h^{AB}T_{AB}.
\label{lagd}
\eeq
We can now follow the procedure discussed in the previous section and
reduce eq.~(\ref{lagd}) into a Lagrangian describing 4-dimensional
fields. We choose the field parametrization in eqs.~(\ref{defg})--(\ref{defp})
and the unitary gauge $\qn_\mu =0$, $\pn_j =0$, $\pn =0$. The meaning
of this gauge choice should be clear by now. It eliminates the
non-physical degrees of freedom absorbed by the massive fields, and it
enables us to work with a Lagrangian with diagonal kinetic terms. In
particular, notice that the shift of $\hn_{\mu \nu}$ proportional to
$\hhn$ in the definition of $\gn_{\mu \nu}$, eq.~(\ref{defg}), is essential
to separate the physical scalar component from the unphysical scalars
contained in the 4-dimensional metric. Without this shift, the kinetic
terms of the Lagrangian would mix $\hhn$ with $\gn_\mu^\mu$ and
$\partial^\mu \partial^\nu \gn_{\mu \nu}$.

In the unitary gauge, the Lagrangian in eq.~(\ref{lagd}) becomes the
sum over the \kk modes
of
\bea
{\cal L} &=& \sum_{{\rm all}~ \vec n} 
 -\frac{1}{2} {G^{(-\vec n)}}^{\mu \nu} (\Box +m^2) 
  \gnv_{\mu \nu}
+\frac{1}{2} {G^{(-\vec n)}}^\mu_\mu (\Box +m^2) {G^{(\vec n)}}^\nu_\nu
-{G^{(-\vec n)}}^{\mu \nu}\partial_\mu 
         \partial_\nu {G^{(\vec n)}}^\lambda_\lambda \nonumber \\ &&
+{G^{(-\vec n)}}^{\mu \nu}\partial_\mu \partial_\lambda 
{G^{(\vec n)}}^\lambda_\nu 
-\frac{1}{4} \left| \partial_\mu \vnv_{\nu j}-\partial_\nu \vnv_{\mu j}
\right|^2 +\frac{m^2}{2}{V^{(-\vec n)}}^{\mu j} \vnv_{\mu j} \nonumber \\ &&
-\frac{1}{2} {S^{(-\vec n)}}^{jk} (\Box +m^2)\snv_{jk} \nonumber 
-\frac{1}{2} H^{(-\vec  n)} (\Box +m^2) \hhnv \nonumber \\ &&
-\frac{1}{\ms}\left[ {G^{(\vec n)}}^{\mu \nu}-\frac{\kappa}{3} 
\eta^{\mu \nu} \hhnv\right]
T_{\mu \nu}. \nonumber
\label{freelag}
\eea
Here $m^2\equiv \n^2$ is the \kk
graviton squared mass and $\kappa$ is given
in eq.~(\ref{cappa}).
The graviton propagator is obtained by inverting the Fourier-transformed
bilinear Lagrangian in eq.~(\ref{freelag}),

\epsfbox{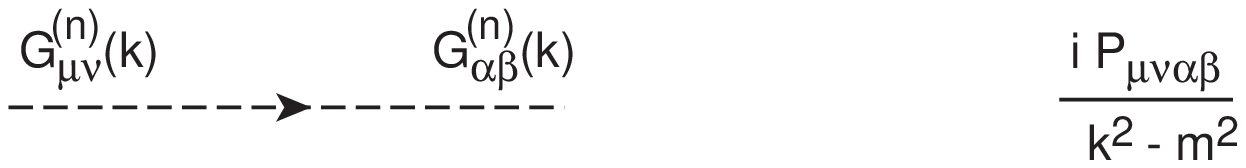}

\bea
P_{\mu\nu\alpha\beta} & = &\frac{1}{2}\left( \eta_{\mu\alpha}\eta_{\nu\beta}
 +\eta_{\mu\beta}\eta_{\nu\alpha}
  -\eta_{\mu\nu}\eta_{\alpha\beta}\right) 
\label{propg}
\\ \nonumber
 & & -\frac{1}{2m^2}\left( 
\eta_{\mu\alpha}{k_\nu k_\beta}
 +\eta_{\nu\beta}{k_\mu k_\alpha}+
  \eta_{\mu\beta}{k_\nu k_\alpha} +\eta_{\nu\alpha}
  {k_\mu k_\beta}\right) \\ \nonumber
 & & +\frac{1}{6}
\left( \eta_{\mu\nu} 
    +\frac{2}{m^2}    k_\mu k_\nu  \right)
\left( \eta_{\alpha\beta} 
    +\frac{2}{m^2}    k_\alpha k_\beta  \right) .
\eea
The spin-sum of the polarization tensors is 
\beq
\sum_s e_{\mu\nu}(k,s)e_{\alpha\beta}(k,s)=
  P_{\mu\nu \alpha\beta}(k) .
\eeq
The tensor $P_{\mu\nu \alpha\beta}(k)$ satisfies conditions 
(\ref{hg2})--(\ref{hg3}), since on mass-shell
\bea
\eta^{\alpha\beta}P_{\mu\nu \alpha\beta}(k) & = & 0, \\
k^\alpha P_{\mu\nu \alpha\beta}(k) & = & 0.
\eea

For our considerations it is also useful to derive the massless
graviton propagator in $D$ dimensions. In order to invert the kinetic term,
we need to add to the Lagrangian in eq.~(\ref{lagd})
a gauge-fixing term, which we choose to be
\beq
{\cal L}=\frac{1}{\xi}C^A C_A ,
~~~C_A
= \partial^B h_{A B} -\frac{1}{2} \partial_A h^B_B .
\eeq
Here $\xi$ is the gauge-fixing parameter, and $\xi =1$ corresponds to the
de Donder gauge often chosen in quantum gravity. The propagator of the
massless graviton is $iP^{(0)}_{ABCD}/k^2$, where
\bea
P^{(0)}_{ABCD} & = &\frac{1}{2}\left( \eta_{AC}\eta_{BD}
 +\eta_{AD}\eta_{BC}\right) -\frac{1}{D-2}
  \eta_{AB}\eta_{CD}
\label{propgd}
\\ \nonumber
 & & +\frac{(\xi -1 )}{2k^2}\left( 
\eta_{AC}{k_B k_D}
 +\eta_{BD}{k_A k_C}+
  \eta_{AD}{k_B k_C} +\eta_{BC}
  {k_A k_D}\right) .
\label{prop0}
\eea

We now turn to 
the graviton interaction Lagrangian, which is given by
\beq
{\cal L} = -\frac{1}{\ms} \gn_{\mu \nu} T^{\mu \nu}.
\label{lint}
\eeq
Here $T_{\mu \nu}$ is the energy-momentum tensor, which is symmetric and
conserved. 
As previously discussed, the
\kk excitations of the graviton have the
same couplings to ordinary fields as their massless zero mode. 

We start by discussing the case of QED coupled to quantum gravity, which
is described by
the Lagrangian 
\beq
{\cal L}=\sqrt{-g} \left( i \bar \psi \gamma^a D_a \psi
-\frac{1}{4}F_{\mu \nu}F^{\mu \nu} \right),
\eeq
\beq
D_a=e^\mu_a\left( \partial_\mu -ieQA_\mu +\frac{1}{2} \sigma^{bc}
e_b^\nu \partial_\mu e_{c \nu} \right) .
\eeq
Here $\sigma^{ab}=(\gamma^a \gamma^b - \gamma^b \gamma^a )/4$, $e^\mu_a$
are the vierbein fields, and $Q$ is the electric charge of the fermion $\psi$.
Greek indices refer to general coordinate transformations and
Latin indices to Lorentz transformations. By varying the Lagrangian
with respect to the vierbein, we obtain
\bea
T_{\mu \nu}&=& \frac{i}{4}\bar \psi (\gamma_\mu \partial_\nu +
\gamma_\nu \partial_\mu )\psi - \frac{i}{4}(\partial_\mu \bar \psi 
\gamma_\nu +\partial_\nu \bar \psi 
\gamma_\mu) \psi +
\nonumber \\ &&
\frac{1}{2} eQ\bar \psi (\gamma_\mu A_\nu
+ \gamma_\nu A_\mu )\psi +F_{\mu \lambda}{F^\lambda}_\nu +\frac{1}{4}
\eta_{\mu \nu} F^{\lambda \rho} F_{\lambda \rho} .
\label{tqed}
\eea
Notice that the trace of the above energy-momentum
tensor is zero on the equations of motion, as a result of the
tree-level conformal invariance of a massless gauge theory.

{}From eq.~(\ref{lint}), we obtain the following Feynman rules:

\epsfbox{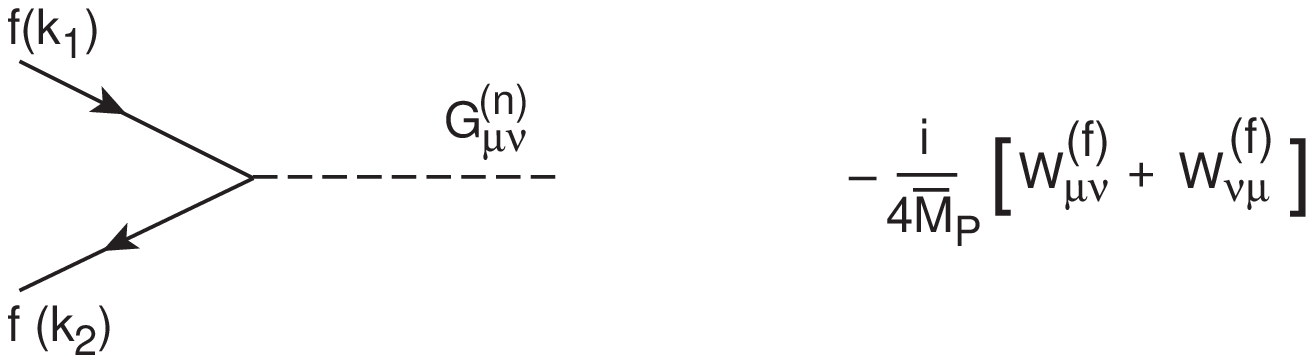}

\beq
W^{(f)}_{\mu \nu}=(k_1+k_2)_\mu \gamma_\nu
\eeq

\epsfbox{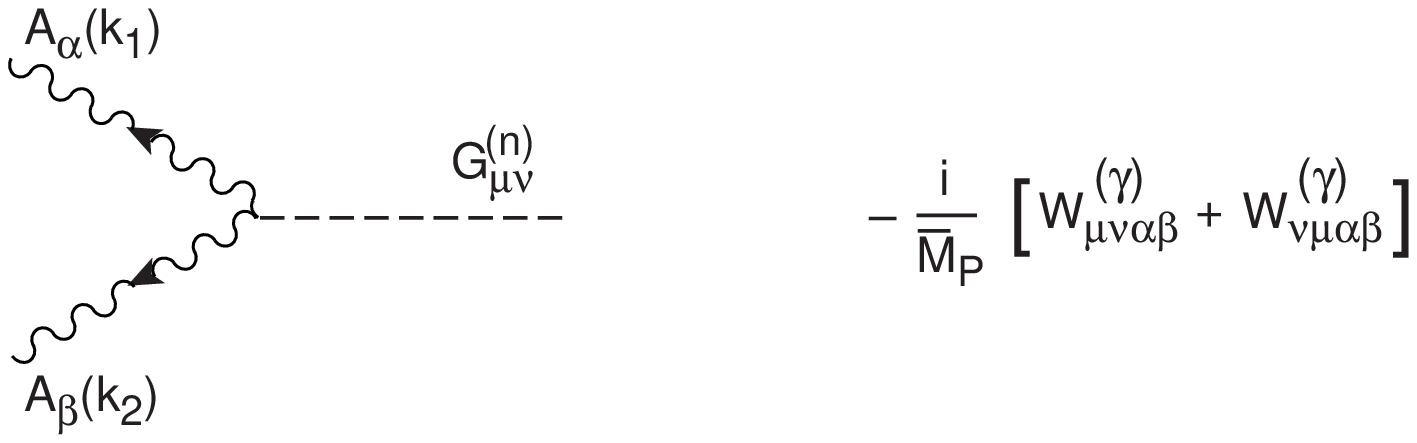}

\bea
W^{(\gamma)}_{\mu \nu\alpha \beta}& = & \frac{1}{2}\eta_{\mu\nu}
  \left( k_{1\beta}k_{2\alpha}-k_1\cdot k_2 \eta_{\alpha\beta} \right)
+ \eta_{\alpha\beta}k_{1\mu}k_{2\nu}\\ \nonumber
& &+\eta_{\mu\alpha}\left( k_1\cdot k_2 \eta_{\nu\beta}-k_{1\beta}k_{2\nu}
\right)-\eta_{\mu\beta}k_{1\nu}k_{2\alpha}
\eea

\epsfbox{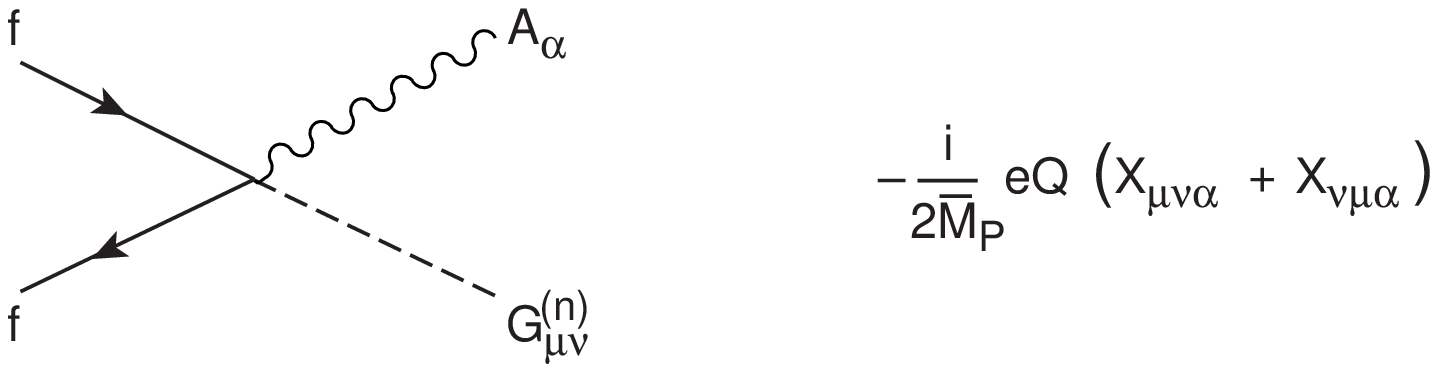}

\beq
X_{\mu \nu \alpha} = \gamma_\mu \eta_{\nu \alpha}.
\eeq
We follow the convention that 
particle momenta (indicated inside parenthesis when necessary) flow along the
arrow directions.

The generalization of these results to the case of QCD coupled to
quantum gravity is straightforward.
The energy-momentum tensor for QCD has the same form as eq.~(\ref{tqed}) 
with the replacements $eQA_\mu \to g_sA_\mu^at^a$, $F_{\mu\nu}
\to G_{\mu \nu}^a= \partial_\mu A^a_\nu - \partial_\nu A^a_\mu
-g_s f^{abc}A_\mu^bA_\nu^c$, where $a,b,c$ are group indices,
$t^a$ are group generators in the fundamental representation, and
$f^{abc}$ are the structure constants. We then obtain the following
Feynman rules:

\epsfbox{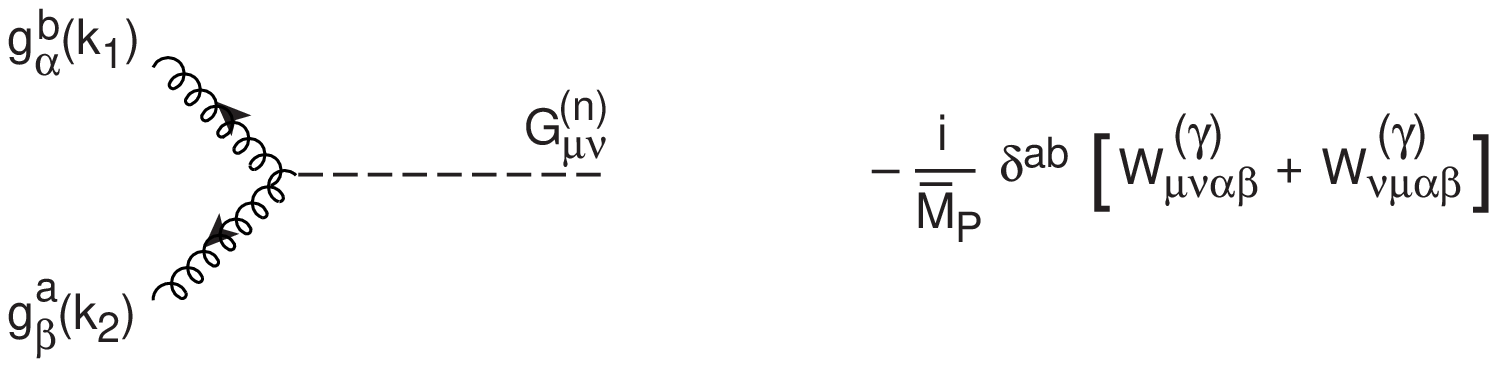}

\epsfbox{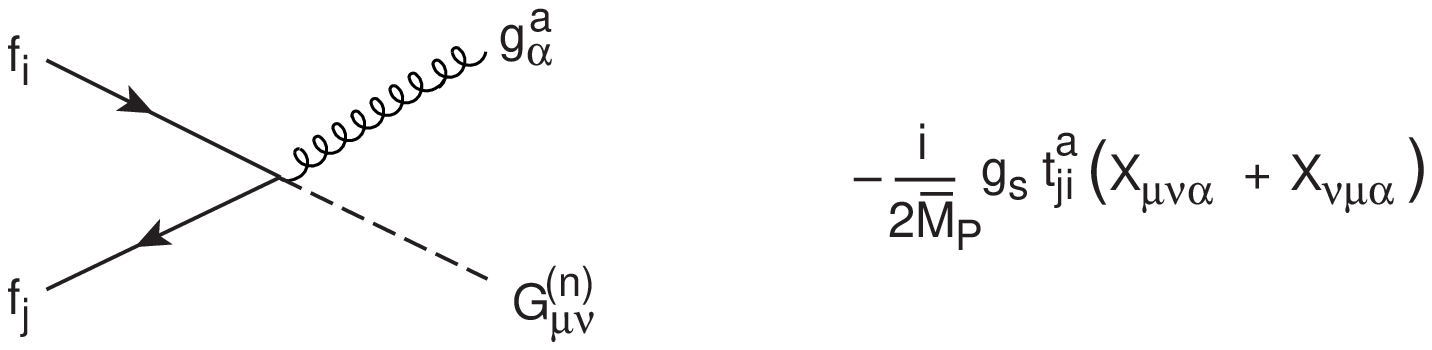}

\hspace*{-0.5cm}
\epsfbox{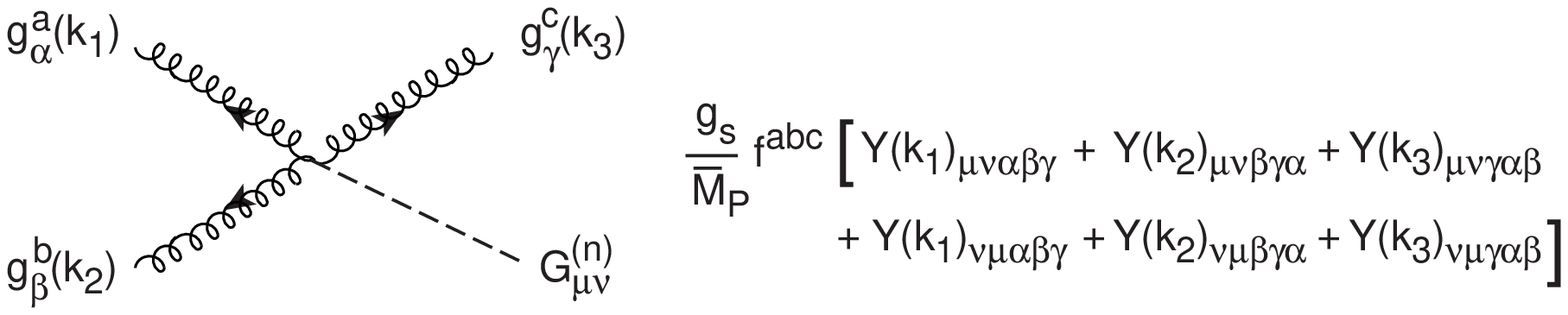}

\bea
Y(k) &=& k_\mu \left( \eta_{\nu \beta} \eta_{\alpha \gamma} 
                    - \eta_{\nu \gamma} \eta_{\alpha \beta} \right)
\nonumber \\ &+&
       k_\beta \left( \eta_{\mu \alpha} \eta_{\nu \gamma}
 -\frac{1}{2}         \eta_{\mu \nu} \eta_{\alpha \gamma} \right)
    - k_\gamma \left( \eta_{\mu \alpha} \eta_{\nu \beta}
 -\frac{1}{2}         \eta_{\mu \nu} \eta_{\alpha \beta} \right) .
\eea
Since we will not make use of the vertex with four gluons and a graviton,
we will not write it down explicitly.

\section{Graviton Production Processes}
\label{seccros}

We now consider the physical processes relevant to collider
experiments, starting with graviton production.
The graviton
\kk modes have masses equal to $|n|/R$, and therefore the different
excitations have mass splittings
\beq
\Delta m \sim \frac{1}{R} = \md \left( \frac{\md}{\ms}\right)^{2/\delta} 
\sim \left( \frac{\md}{{\rm TeV}}\right)^{\frac{\delta +2}{2}}~
10^{\frac{12\delta -31}{\delta}}{\rm eV}.
\eeq
Here we have used the relation between the compactification radius $R$
and the effective Planck mass of the $D$-dimensional theory $\md$, 
eq.~(\ref{planc}).
If we take $\md =$ 1 TeV and $\delta =4,6,8$, then $\Delta m$ is equal
to 20 keV, 7 MeV, and 0.1 GeV, respectively. Only for a large number
of extra dimensions does the mass splitting become comparable with the
experimental energy resolution. However, for large $\delta$, only a small
number of \kk modes can be produced and the total 
cross-section is negligible. We are interested in the case in which $\delta$ 
is not too large (say $\delta \lsim 6$), where the enormous number of
accessible \kk modes can compensate the $1/\ms^2$ factor in
the scattering amplitude. 

For experimental applications, it is convenient to express the
graviton-production rate in terms of inclusive cross-sections, where
the contributions of the different \kk modes have been summed up.
For not too large $\delta$, the mass splitting
$\Delta m$ is so small that the sum over the
different \kk states can be replaced by a continuous integration.
The number of modes with \kk index between $|n|$ and $|n|+dn$
is given by
\beq
dN=S_{\delta -1} |n|^{\delta -1} dn,~~~~~S_{\delta 
-1}=\frac{2\pi^{\delta /2}}{\Gamma (\delta /2 )},
\label{dn}
\eeq
where $S_{\delta -1}$ is the surface of a unit-radius 
sphere in $\delta$ 
dimensions\footnote{For $\delta=2n$ and $n$ integer, 
we find
$S_{\delta -1}=2\pi^n/(n-1)!$ and, for $\delta=2n+1$, we find 
$S_{\delta -1}=2\pi^n/\Pi_{k=0}^{n-1} (k+\frac{1}{2})$.}.
Using eq.~(\ref{planc}) and the expression
$m=|n|/R$ for the \kk excitation mass, we can rewrite the
measure in eq.~(\ref{dn}) as
\beq
dN=S_{\delta -1} ~\frac{\ms^2}{\md^{2+\delta}}~ m^{\delta -1} ~dm .
\label{dnn}
\eeq
Hence we express the differential cross-section for inclusive
graviton production
as
\beq
\frac{d^2\sigma}{dt ~dm} = S_{\delta -1}~ \frac{\ms^2}{\md^{2+\delta}}~
m^{\delta -1}~ \frac{d\sigma_m}{dt},
\label{sigf}
\eeq
where $d\sigma_m/dt$ is the differential cross-section for producing a
single \kk graviton of mass $m$. 
Since the graviton interaction vertex is suppressed by $1/\ms$ (see
sect.~\ref{secfey}) we can anticipate that $\sigma_m\propto \ms^{-2}$, and
the factor $\ms^2$ appearing from the phase-space summation exactly cancels
the Planck mass  dependence in eq.~(\ref{sigf}).

The same result can also be obtained from the point of view of the
$D$-dimensional theory. Here, the canonically normalized graviton 
$h_{AB}$ has a coupling to matter ${\cal L}_D=-T^{AB} h_{AB}/
\bar M_D^{1+\delta/2}$.
The graviton phase space is
\beq
d\Phi_G =\frac{d^{D-1} k_D}{(2\pi)^{D-1} 2\sqrt{k_D \cdot k_D}}=
{d^\delta k_T\over (2\pi)^\delta}\times \frac{d^3 k}{(2\pi)^3 2 
\sqrt{ {\vec{k}}^2 +m^2}},
\label{meas}
\eeq
where we have decomposed $k_D$ in the components $k_T$ and $k$ 
respectively orthogonal and parallel to the brane. Graviton emission from the
brane is, in lowest order, independent of the orientation of $k_T$
so we can trivially integrate on it: 
$d^\delta k_T= S_{\delta-1}m^{\delta-1}d m$. The cross section for
an initial brane state $|p_1,p_2\rangle$ to go in a final brane 
state $|f\rangle$
plus a bulk graviton $|G\rangle$ is then
\beq
d\sigma ={S_{\delta-1}m^{\delta-1}dm\over M_D^{2+\delta}}
  \bigl |\langle f,G|T^{\mu\nu}
 h_{\mu\nu}|p_1,p_2\rangle\bigr |^2 (2\pi)^4\delta^4(P_{i}-P_f) {d\Phi_f\over 
F(p_1,p_2)},
\eeq
where $d\Phi_f$ is the brane final state phase space, and $F(p_1,p_2)$
is just the usual flux factor for two particle collision. This equation 
agrees with eq.~(\ref{sigf}).


We now give the differential cross-sections for producing a \kk
graviton of mass $m$ in processes
which are relevant to the high-energy collider experiments discussed in
the sect.~\ref{seccol}. 
The cross-section for producing a graviton and a photon
in a fermion-antifermion collision is
\beq
\frac{d\sigma_m}{dt}(f\bar f \to \gamma G) = \frac{\alpha Q_f^2}{16N_f}~
\frac{1}{s\ms^2}~F_1 (t/s,m^2/s).
\label{sigfer}
\eeq
Here $Q_f$ and $N_f$ are the electric charge and number of colours of
the fermion $f$, and $F_1$ is provided in the appendix.
The cross-sections for the parton processes relevant to graviton plus
jet production in hadron collisions
are
\beq
\frac{d\sigma_m}{dt}(q\bar q \to g G) = \frac{\alpha_s}{36}~
\frac{1}{s\ms^2}~F_1 (t/s,m^2/s)
\label{sigqq}
\eeq
\beq
\frac{d\sigma_m}{dt}(qg \to q G) = \frac{\alpha_s}{96}~
\frac{1}{s\ms^2}~F_2 (t/s,m^2/s)
\label{sigqg}
\eeq
\beq
\frac{d\sigma_m}{dt}(gg \to g G) = \frac{3\alpha_s}{16}~
\frac{1}{s\ms^2}~F_3 (t/s,m^2/s)
\label{siggg}
\eeq
The Mandelstam variable $t$ in eq.~(\ref{sigqg}) is defined as $t=(p_q -p_G)^2$.
The functions $F_2$ and $F_3$ are provided in the appendix.

We want to stress again that our calculation of graviton emission is
based on an effective low-energy theory, valid below the scale $\md$. 
Of course, we cannot determine the precise scale at which our effective
low-energy theory breaks down.  Nevertheless, we can use naive dimensional
analysis to estimate the energy scale at which perturbation theory
breaks down. If we analyse graviton loop corrections in $D$ dimensions,
we observe that the expansion parameter is \cite{donoghue}
\beq
\frac{S_{D-1}}{2(2\pi)^D}\left( \frac{E}{\md}\right)^{D-2}.
\eeq
Here, $S_{D-1}$ is the surface of the $D$-sphere given in eq.~(\ref{dn}),
and $E$ is the relevant energy of the process. Requiring that the expansion
parameter is less than unity, we obtain an estimate of
the maximum energy at which
we can trust a perturbative expansion in the effective theory
\beq
E_{\rm max} = \left[ \Gamma (2+\delta /2)\right]^{\frac{1}{2+\delta}}
\left( 4\pi \right)^{\frac{4+\delta}{4+2\delta}} \md .
\eeq
Numerically, one finds always that $E_{\rm max}>7.2M_D$.
Although perturbativity can be trusted up to $E_{\rm max}$, new quantum-gravity
effects may appear much sooner, as mentioned in the introduction.

\section{Virtual Graviton Exchange}
\label{secvir}

We now want to study the effect of a single virtual-graviton exchange
at tree-level 
in scattering processes. For simplicity
we consider the case of pure $s$-channel exchange, but the discussion
of the $t$- and $u$-channel exchange is completely analogous.
The scattering amplitude in momentum space of the graviton-mediated
process is
\beq
{\cal A} = \frac{1}{\ms^2}  \sum_n \left[ T_{\mu \nu} 
\frac{P^{\mu \nu\alpha\beta}}{s-m^2}
T_{\alpha \beta} + \left(\frac{\kappa}{3}\right)^2 
\frac{T_\mu^\mu T_\nu^\nu}{s-m^2}\right]
={\cal S}(s)~{\cal T}
\label{ampl}
\eeq
\beq
{\cal S}(s)\equiv \frac{1}{\ms^2}  \sum_n \frac{1}{s-m^2} 
\eeq
\beq
{\cal T}\equiv T_{\mu \nu}T^{\mu \nu}-\frac{1}{\delta +2}T^\mu_\mu T^\nu_\nu .
\label{top}
\eeq
Here $T_{\mu \nu}$ is the energy-momentum tensor, $k$ is the transferred
momentum, $\kappa$ is defined in eq.~(\ref{cappa}), and
$P^{\mu \nu\alpha\beta}$
in eq.~(\ref{propg}). The two terms in eq.~(\ref{ampl}) correspond to
the exchange of the graviton $\gn_{\mu \nu}$ and the scalar $\hhn$. The
same result can be obtained by exchanging a massless $D$-dimensional
graviton with the propagator in eq.~(\ref{prop0}) and with coupling to
the energy-momentum tensor given in eq.~(\ref{tensor}).
In eq.~(\ref{ampl}) 
$\sum_n$ represents the sum over all \kk modes, which has to be
performed at the amplitude level.
Since the operator ${\cal T}$ does not depend on the \kk index, we can
can perform the sum $\sum_n$
without specifying the particular physical scattering
process under consideration. It is convenient to convert the sum into 
an integral which can be evaluated using dimensional
regularization:
\beq
{\cal S}(s)=
\frac{1}{\md^{2+\delta}}\int d^\delta q_T \frac{1}{s-q_T^2}
= \pi^{\frac{\delta}{2}} \Gamma (1-\delta /2)\left( -\frac{s}{\md^2}
\right)^{\frac{\delta}{2}-1}.
\label{scal}
\eeq
Here we used $m^2=q_T^2$, with $q_T$ the graviton momentum 
orthogonal to the brane.
To further simplify eq.~(\ref{scal}), we need to distinguish the cases
in which $\delta$ is even or odd. For even $\delta$ ($\delta \geq 4$), we find
\beq
{\cal S}(s) = -\frac{1}{\md^{2+\delta}}\frac{S_{\delta -1}}{2}
\left[ \left(i\pi +\ln \frac{s}{\mu^2}\right) s^{\frac{\delta -2}{2}}
+\sum_{k=1}^{(\delta -2)/2}
c_k ~\Lambda^{\delta -2k}~s^{k-1}\right] 
\label{seven}
\eeq
and for odd $\delta$ ($\delta \geq 3$),
\beq
{\cal S}(s) = -\frac{1}{\md^{2+\delta}}\frac{S_{\delta -1}}{2}
\left[ i\pi \sqrt{s}~ s^{\frac{\delta -3}{2}}+\sum_{k=1}^{(\delta -1)/2}
c_k ~\Lambda^{\delta -2k}~s^{k-1}\right] .
\label{sodd}
\eeq
Here $S_{\delta -1}$ is defined in eq.~(\ref{dn}), $\mu$ is the subtraction
mass, $\Lambda$ is an
ultraviolet cutoff, and $c_k$ are unknown coefficients. The terms
proportional to $\Lambda$ describe divergent terms not computable in
the effective theory. The presence of ultraviolet divergences in
tree-level processes is related to existence of an infinite tower
of \kk modes.  In the case $\delta =2$, there are no power divergences
and ${\cal S}(s)$ is given by eq.~(\ref{seven}) with the summation over
$k$ omitted.

The non-analytic pieces proportional to $\ln (-s)$ and $\sqrt {-s}$
in ${\cal S}(s)$ are determined by low-energy physics only. This is because 
they have branch-cut singularities -- the imaginary parts in eqs.~(\ref{seven}),
(\ref{sodd}) -- corresponding to real graviton emission. Therefore
this dependence is uniquely fixed as it  cannot
be affected by the introduction of local counterterms~\cite{donoghue}. 
Notice that these effects are
the analogue of the familiar ``running" of amplitudes in quantum
field theory, although here we are apparently 
dealing with a tree-level calculation.

Unfortunately, ${\cal S} (s)$ is dominated by the ultraviolet contributions,
which can be computed only with some knowledge of the underlying 
quantum-gravity theory. The infrared contributions could be experimentally
isolated only if the coefficients $c_k$ turn out to be small (because of
some ``miraculous" cancellation of divergences in the fundamental
theory), or with very precise measurements of the energy dependence
or angular dependence (in case of $t$-channel graviton exchange) of
the scattering process. 
The case $\delta =2$ is particularly interesting, since there is only
a logarithmic dependence on the cutoff. However,
for phenomenological applications, we will adopt
the most plausible assumption
that ${\cal S} (s)$ is 
dominated by the lowest-dimensional local operator,
and that
the amplitude
of the physical process is described by
\beq
{\cal A} = \frac{S_{\delta -1}}{2}~c_1~
\frac{\Lambda^{\delta -2}}{\md^{\delta +2}}~{\cal T}
\equiv \frac{4\pi}{\Lambda_T^4}~{\cal T} ~~~~{\rm for}~\delta >2.
\label{amplt}
\eeq
Here $\Lambda$ is the unknown cutoff energy, presumably of order $\md$.
In the case of string theory, the effective cutoff could appear at
a scale lower than $\md$, giving a suppression of the amplitude.

{}From eq.~(\ref{amplt}), we can compute the cross-sections for various
processes relevant to collider experiments. For instance, the cross-sections 
for processes with two photons in the final state are
\beq
\frac{d\sigma}{dt}(f\bar f \to \gamma\gamma )=\frac{2\pi}{N_f s^2}
\left[ \alpha Q_f^2 G_1(t/s)+\frac{2s^2}{\Lambda_T^4} G_2(t/s)\right]^2
\eeq
\beq
\frac{d\sigma}{dt}(gg\to \gamma\gamma )=\frac{\pi}{16}\frac{s^2}{\Lambda_T^8}
G_3(t/s),
\eeq
and the cross-section for the fermion scattering process $e^+ e^- \to f\bar f$
(with $f\neq \nu_e$) is
\bea
&&\frac{d\sigma}{dt}(e^+e^- \to f\bar f)=
\frac{d\sigma}{dt}(e^+e^- \to f\bar f)_{\rm SM}
+\frac{N_f\pi}{32}\frac{s^2}{\Lambda_T^8}G_4(t/s)
\nonumber \nonumber \\
 && -\frac{N_f\alpha \pi}{2\Lambda_T^4}\left\{ Q_e Q_{f} G_5(t/s)
+\frac{1}{\sin^2 2\theta_W}\frac{s}{s-M_Z^2}
\left[ v_ev_{f}G_5(t/s)+a_ea_{f}G_6(t/s)\right] \right\} \nonumber \\
&& -\frac{\alpha \pi}{2\Lambda_T^4}\delta_{ef}\left\{
Q_e^2 G_7(t/s)
+\frac{s}{\sin^2 2\theta_W}\frac{v^2_e+a^2_e}{s-M_Z^2}G_8(t/s)\right. \nonumber \\
&& \left. +\frac{1}{\sin^2 2\theta_W}
 \frac{s}{t-M_Z^2}\left[ v^2_eG_{9}(t/s)+a^2_eG_{10}(t/s)\right]\right\}
+\frac{\delta_{ef}\pi}{32} \frac{s^2}{\Lambda_T^8} G_{11}(t/s),
\eea
where $v_f=T_f-2Q_f\sin^2\theta_W$, $a_f=T_f$ and $t=(p_{e^-}-p_f)^2$. 
The symbol $\delta_{ef}$
is equal to 1 if the final-state fermion is an electron ($f=e$), and is
equal to 0 otherwise. The $G_i(x)$ functions
are given in the appendix.  
Collider tests of the existence of the operator ${\cal T}$ will be
discussed in sect.~\ref{secop}, where we study the $f\bar f\to \gamma\gamma$
case as an example.
 
Before concluding this section, we use
eqs.~(\ref{seven})--(\ref{sodd}) to derive a limit
on the applicability of the effective theory from unitarity arguments.
Using the Wigner-Eckart
theorem, we can decompose the transition amplitude between an initial
state $i$ and a final state $f$ into helicity amplitudes 
\beq
{\cal A}_{fi} = 8\pi \sum_J (2J+1){\cal D}_{fi}^J (\theta)
\langle f|T^J|i\rangle .
\eeq
Here ${\cal D}$ are the Wigner functions which depend on the scattering
angle $\theta$.
Let us now consider the scattering process between two pairs of different
fermions $f\bar f \to G \to f'\bar f'$, mediated by 
graviton exchange in the $s$-channel. 
For an initial and final state of helicity one, we obtain
\beq
\langle f_{\lambda =1}|T^{J=2}|i_{\lambda =1}\rangle =\frac{s^2}{160\pi}~
{\cal S}(s).
\label{lang}
\eeq
By unitarity it must be $|T_{fi}^{J=2}|<1$, and in particular
$|{\rm Im}T_{fi}^{J=2}|<1$. Therefore, using eqs.~(\ref{seven}) and (\ref{sodd}),
we find that
unitarity is violated unless
\beq
\sqrt{s}<\left[ 160\pi^{-\delta/2}\Gamma (\delta /2)
\right]^{\frac{1}{2+\delta}}\md .
\eeq
Here, $\sqrt{s_{\rm max}}/M_D$ is greater than $1.4$ for $\delta \leq 5$.

\section{Graviton Production and Collider Experiments}
\label{seccol}

We will now discuss the possibility of studying graviton production at
high-energy colliders. Gravitons couple to matter only gravitationally,
but the $1/\ms^2$ suppression present in their production cross-section can be
compensated by the large multiplicity of the \kk modes or, in other
words, by the $D$-dimensional phase-space factor. However the
$1/\ms^2$ suppression in the graviton decay rate into ordinary matter
is not compensated by phase space
and
therefore its lifetime is $\tau_G \sim \ms^2/m^3\sim
({\rm TeV}/m)^3 ~10^3$ seconds. The $1/\ms^2$ suppression factor can be also
interpreted as the small probability that a graviton propagating in the 
$D$-dimensional space crosses the 3-dimensional brane. 

For experimental
purposes, this means that the \kk graviton behaves like a massive,
non-interacting, stable particle and its collider signature is 
imbalance in final state momenta and missing mass. 
Since the relevant observables for graviton production are described only by
inclusive cross-sections, the graviton has a continuous distribution in
mass, described by eq.~(\ref{sigf}). This mass distribution corresponds
to the probability of emitting gravitons with different momenta
in the extra dimensions. Notice that this is a peculiarity of the graviton
signal with respect to other new-physics processes.  For instance,
supersymmetry
with conserved R-parity also can yield an excess of
missing energy events,
but these correspond to a fixed invisible-particle mass.
Signals with topologies similar to those
discussed in the following can also be encountered in phenomenological
scenarios with ultralight gravitinos~\cite{zwi}. 
At any rate, graviton production leads to energy and angular distributions 
that are in general distinct from those corresponding to other
new-physics processes. Moreover, in the case of supersymmetry, the
missing-energy signal is always accompanied by a variety of leptons, photons,
and hadronic activity coming from the decay of heavier particles. For
graviton-production in the perturbative regime, 
each extra particle in the final state
is associated with an extra suppression factor.

The emission of a single graviton in the extra dimensions violates 
momentum conservation along the directions transverse to the brane. 
This is not surprising, since translational invariance in the $D$-dimensional
space is broken by the presence of the brane. In other words, the brane
can radiate gravitons into the extra dimensions conserving the total
energy (since time invariance is preserved), but absorbing any arbitrary
transverse momentum smaller than the energy tension. From a 4-dimensional
point of view, energy and momentum are conserved, but the \kk gravitons
can have any arbitrary mass smaller than about $\md$, the approximate 
cutoff on the validity of our description.

\subsection{$e^+e^-$ and Muon Colliders}
\label{sectee}

We start our analysis by studying future high-energy $e^+e^-$
and $\mu^+ \mu^-$ colliders. Since both cases require an analogous discussion,
for simplicity we will just refer to $e^+e^-$ collisions.
We will focus on the graviton-production process $e^+e^-\to \gamma G$,
but similar considerations hold for $e^+e^-\to Z G$. Using
eqs.~(\ref{sigf}) and (\ref{sigfer}), the corresponding
differential cross-section is
\beq
\frac{d^2\sigma}{dx_\gamma d\cos \theta}(e^+e^- \to \gamma G)
=\frac{\alpha}{64}~S_{\delta -1}~ 
\left( \frac{\sqrt{s}}{\md}\right)^{\delta +2}
\frac{1}{s}f(x_\gamma, \cos\theta)
\label{sezgg}
\eeq
\bea
f(x,y)&=&x(1-x)^{\frac{\delta}{2} -1} ~F_1 \left( -\frac{x}{2}(1-y),1-x \right)
\nonumber \\
&=&\frac{2(1-x)^{\frac{\delta}{2} -1}}{x(1-y^2)}
\left[ (2-x)^2(1-x+x^2)-3y^2x^2(1-x)-y^4x^4 \right] .
\label{funz}
%
\eea
Here $x_\gamma =2E_\gamma /\sqrt{s}$, $E_\gamma$ is the photon energy,
and $\theta$ is the angle between the photon and beam directions.
Although we are considering a two-body process, the differential 
cross-section depends on two kinematic variables, because of the continuous
distribution in the graviton mass $m=\sqrt{s(1-x_\gamma)}$. 
The differential cross-section in
eq.~(\ref{sezgg}) diverges as $x_\gamma \to 0$ or $\cos^2 \theta \to 1$.
This is caused by the collinear divergence originating
{}from the massless fermion exchange 
in the $t$ channel. Notice that, for $\delta >2$,
the factor $(1-x)^{\frac{\delta}{2} -1}$
in eq.~(\ref{funz}) tilts the photon energy spectrum towards small values
of $E_\gamma$. The origin of this effect is the much wider graviton
phase space available at large values of $m$ (see eq.~(\ref{sigf})).

The Standard Model background comes predominantly from the process
$e^+e^-\to \gamma \nu \bar{\nu}$. The peak contribution from 
$e^+e^-\to \gamma Z$ can be eliminated by excluding the 
photon-energy region around $E_\gamma = (s-M_Z^2)/(2\sqrt{s})$. On the other
hand, the remaining continuous distribution in $E_\gamma$ from 
$e^+e^-\to \gamma \nu \bar{\nu}$ represents a significant background.
Other background
contributions, {\it e.g.} from $e^+e^-\to \gamma
(e^+e^-)$ or $e^+e^-\to \gamma (\gamma )$, are not important in the region
of large photon transverse energy we will consider below.

In fig.~\ref{fige1} we show the total cross-section for $E_{T,\gamma}
\equiv E_\gamma
\sin \theta > E_{T,\gamma}^{\rm min}$ and $E_\gamma <450$ GeV at a 
hypothetical future collider with $\sqrt{s}=1$
TeV. 
\jfig{fige1}{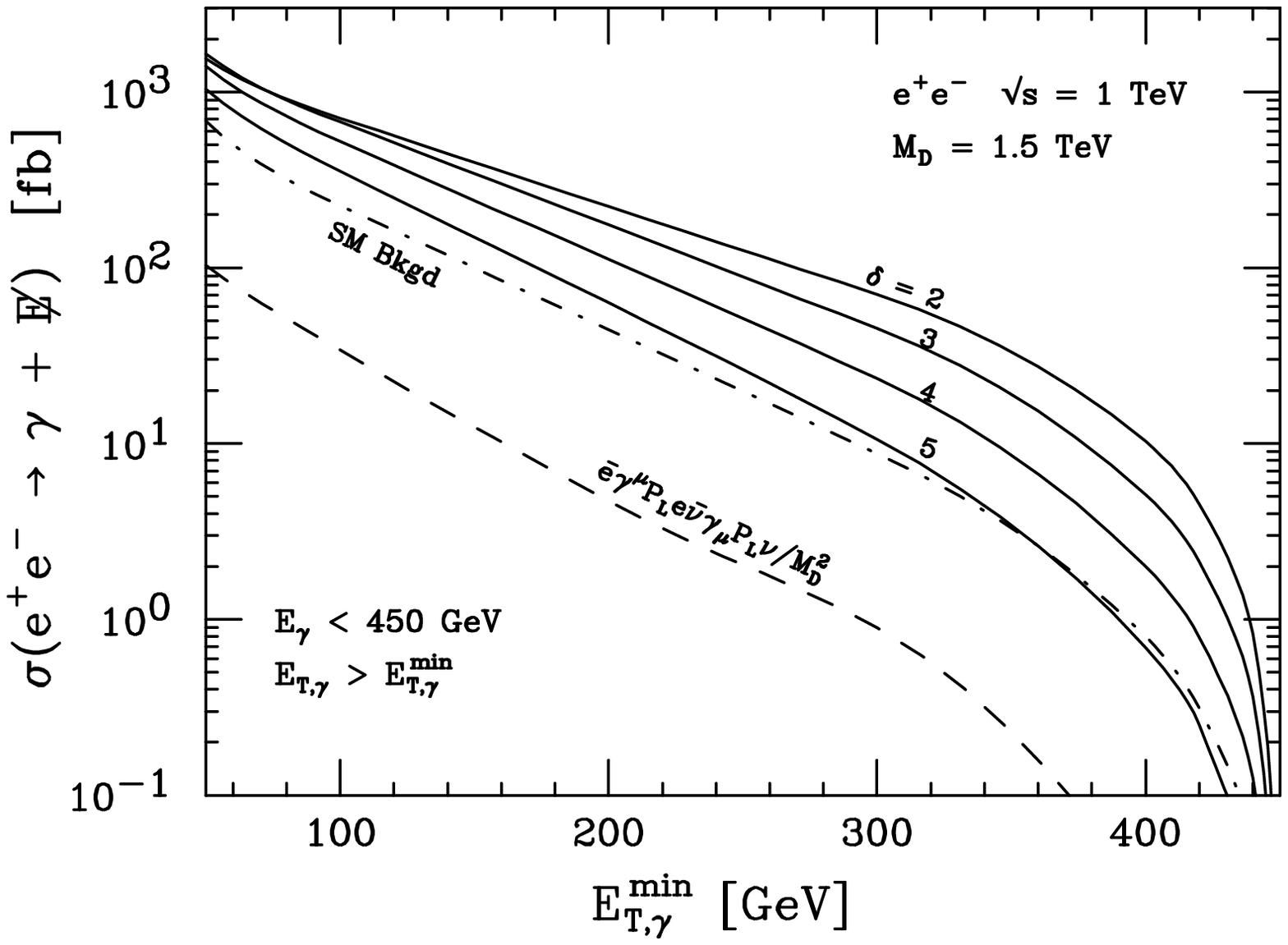}{Total $\gamma + {\rm nothing}$ cross-section
at an $e^+e^-$ collider for $\sqrt{s}=1\tev$ with
$E_{T,\gamma}>E_{T,\gamma}^{\rm min}$.  The dash-dotted line represents
the background, and the solid lines represent the signal for
various numbers of extra dimensions and $M_D=1.5\tev$.
To eliminate the background contribution
from $\gamma Z\to \gamma \bar \nu\nu$ we have required $E_\gamma < 450\gev$
for both the signal and the background.  The dashed line is the Standard
Model
background subtracted signal from a representative dimension-6 operator.}
The signal is plotted for a value of the $D$-dimensional Planck
scale $M_D=1.5$ TeV, and different numbers of extra 
dimensions $\delta$.
The cut in $E_\gamma$ is chosen to exclude the background contribution
{}from the $Z$ peak. The background shown in fig.~\ref{fige1} has been
computed using the program {\tt COMPHEP}~\cite{comphep}. Since the signal has
been calculated at the leading order, for consistency the background is
computed with the same approximation. Nevertheless, we have compared 
the background calculation
to {\tt NUNUGPV}~\cite{nunu}, which includes 
higher order QED radiative corrections, and
found agreement to within 10\% at this energy.

Notice that signal and background rates have rather similar
behaviours with $E_{T,\gamma}^{\rm min}$. This is because the enhanced
sphericity of the signal with respect to the background is compensated
by the phase-space preference to produce heavier \kk gravitons and
therefore softer photons. The tendency towards heavier gravitons grows
with $\delta$ (see eq.~(\ref{sigf})), as apparent from the steeper decrease
of the curves with larger $\delta$ in 
fig.~\ref{fige1}.

Next we fix $E_{T,\gamma}^{\rm min}=300$ GeV and show in fig.~\ref{fige2}
the signal rate as a function of $\md$, and compare it to the Standard
Model background. 
\jfig{fige2}{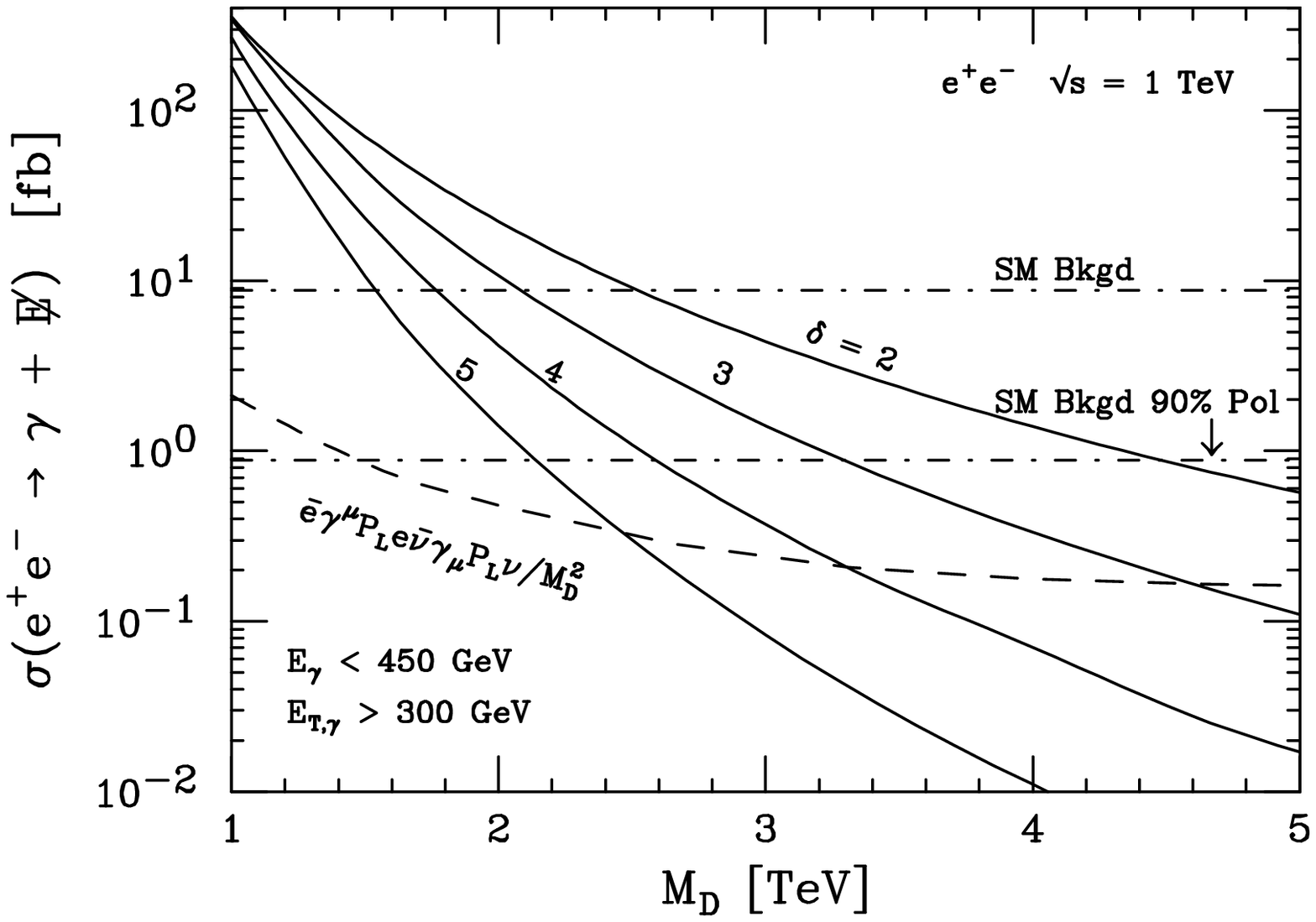}{Total $e^+e^-\to \gamma + {\rm nothing}$ cross-section
at a $1\tev$ centre-of-mass energy $e^+e^-$ collider.  The signal from
graviton production is presented as solid lines for various
numbers of extra dimension ($\delta =2,3,4,5$).  
The Standard Model background for
unpolarized beams is given by the upper dash-dotted line, and the background
with $90\%$ polarization is given by the lower dash-dotted line.  The signal
and background are computed with the requirement $E_\gamma < 450\gev$
in order to eliminate the $\gamma Z\to \gamma\bar\nu\nu$ 
contribution to the background. The dashed line is the Standard Model 
background subtracted signal from a representative dimension-6 operator.}
The dependence of the cross-section on $\md$
is just given by the scaling relation $\sigma \propto 1/\md^{2+\delta}$.
To extract the range of $\md$ which can be probed
at a future $e^+e^-$ collider with $\sqrt{s}=1$ TeV and an integrated
luminosity ${\cal L}= 200$ fb$^{-1}$, we require 
\beq
\sigma_{\rm Signal} > \frac{5\sqrt{ \sigma_{\rm Bkgd}{\cal L}}}{\cal L}
=1.0~{\rm fb}.
\eeq
The results for the corresponding maximum values of $\md$ 
are given in table~1. 
We want to stress that we have not tried to optimize our cuts, and therefore
our results are just indicative of the possibilities. 
Dedicated analyses including 
higher-order corrections, improved cut choices, and detector simulations
are not within the scope of this paper, although they would certainly
increase the region of predicted discovery reach in $\md$.
Notice that the sensitivity
range of $\md$ for colliders with different centre-of-mass energies can
be obtained simply by rescaling fig.~\ref{fige2}, since the variable
$\md$ always appears in the cross-section in the combination $\md
/\sqrt{s}$. 

\begin{table}
\centering
\begin{tabular}{ccc}
\hline\hline
   & Max $\md$  & Max $\md$  \\
$\delta$    & sensitivity  & sensitivity \\
   & $P=$ 0\%    & $P=$ 90\% \\
\hline
2 & 4.1 TeV & 5.7 TeV \\
3 & 3.1~~~~~~  & 4.0~~~~~~  \\
4 & 2.5~~~~~~  & 3.0~~~~~~  \\
5 & 2.0~~~~~~  & 2.4~~~~~~  \\
\hline\hline
\end{tabular}
\caption{{ Maximum $\md$ sensitivity which can be reached
by studying the final state $\gamma +\Emiss$ at an $e^+e^-$ collider
with $\sqrt{s}=1$~TeV, integrated luminosity ${\cal L}=200$~fb$^{-1}$,
and beam polarization $P=$~0\% or 90{\%}. The bounds have been obtained
by requiring $\sigma_{\rm Signal}>1.0$~fb with the acceptance cuts
$E_\gamma <450$~GeV and $E_{T,\gamma}>300$~GeV.
}}
\end{table}

Stringent bounds on $\md$ come from the requirement that graviton
emission does not rapidly cool SN1987A, preventing the occurrence
of the observed neutrino flux. This bound has been estimated in 
ref.~\cite{dim} to be about $\md \gsim 10^{\frac{15-4.5\delta}{\delta 
+2}}$~TeV, {\it i.e.} 30~TeV for $\delta =2$ and 2~TeV for $\delta =3$.
Therefore, the astrophysical argument excludes observable signals
for $\delta =2$, limits the available region for $\delta =3$, and
is insignificant for $\delta >3$. Nevertheless, even for $\delta =2$,
it is still interesting to have an independent laboratory test.
 
The ability to observe the signal is limited by the background which,
with our cut on $E_\gamma < 450\gev$, 
is primarily coming from processes involving
virtual $W$ exchange. Therefore, with the use of polarized beams, the
background can be significantly reduced without affecting the
signal, which is parity invariant. In fig.~\ref{fige2} and table~1
we show the effect of considering
collisions between
an unpolarized positron beam with an electron beam with polarization
$P=90$\% ($P=100$\% for fully right-handed electrons). 

In contrast with the case of $e^+e^-$ colliders, in muon colliders
it appears arduous to
obtain significant polarizations. However,
a peculiarity of the muon collider is the possibility of very precise
beam-energy resolution, due to small initial-state radiation and
bremsstrahlung. This is useful in the search for graviton
emission, since it allows precise measurements of the rapid rise of
the cross-section with $\sqrt{s}$. Such measurements give direct
information on the number of extra dimensions $\delta$ and on the onset
of quantum gravity.

Limits on $M_D$ can be obtained from LEP2 as well.
To estimate the sensitivity to $M_D$,
we calculate the background and signal integrated
over 
\bea
\label{lep2cuts}
10\gev <E_{\gamma}<(s-M_Z^2)/(2\sqrt{s}) -5\gev~~{\rm and}~~
\theta_\gamma > 10^o .
\eea
The upper-limit cut 
on $E_\gamma$
is intended to reduce the $\gamma Z\to \gamma\bar\nu\nu$ background.
The $\gamma\bar\nu\nu$ cross-section with initial state radiative corrections 
is known to be almost a factor of two larger than the tree-level result
at beam energies near $M_Z$~\cite{nunu,was}.  We calculate the background using
{\tt NUNUGPV}~\cite{nunu} which includes these QED radiative corrections.

Table~2 lists the maximum $M_D$ sensitivity which can be reached
at LEP2 from graviton production processes.
We show results for two collider options:  the recently completed
run with $\sqrt{s}=190\gev$ and integrated luminosity
${\cal L}=4\times 150\xpb^{-1}$, and the near-future run with
perhaps $\sqrt{s}=200\gev$ and integrated luminosity
${\cal L}=4\times 500\xpb^{-1}$.  
We define the statistically significant discovery to be
\beq
\sigma_{\rm Signal} > 5\frac{\sqrt{\sigma_{\rm Bkgd}{\cal L}}}{{\cal L}},
\eeq
which is equal to $0.32\xpb$ and $0.17\xpb$ for the $\sqrt{s}=190$ and
200~GeV option, respectively.
The relative fraction
of background to signal is well within systematic errors, and so detecting
a signal is statistically limited.  With more luminosity one
could probe higher values of $M_D$.

\begin{table}
\centering
\begin{tabular}{ccc}
\hline\hline
         & Max $\md$  & Max $\md$ \\
$\delta$    & sensitivity  & sensitivity   \\
 & $\sqrt{s} =190\gev$ & $\sqrt{s} =200\gev$ \\
   & ${\cal L}=4\times 150\xpb^{-1}$    &  ${\cal L}=4\times 500\xpb^{-1}$ \\
\hline
2 & 1100 GeV & 1300 GeV \\
3 & 850~~~~~  & 1000~~~~~~ \\
4 & 700~~~~~ & 800~~~~~ \\
5 & 600~~~~~ & 650~~~~~ \\
\hline\hline
\end{tabular}
\caption{{Maximum $\md$ sensitivity which can be reached
by studying the final state $\gamma +\Emiss$ at LEP2
with $\sqrt{s}=190$~GeV and integrated luminosity 
${\cal L}=4\times 150\xpb^{-1}$, or $\sqrt{s}=200\gev$ and integrated
luminosity ${\cal L}=4\times 500\xpb^{-1}$.}}
\end{table}

Here we have compared the signal of graviton production with the
Standard Model background. However, the unknown physics at scales
larger than $\md$ can produce other phenomena which can provide
an unexpected source of background to the graviton signal. The effect
of the ultraviolet physics can be absorbed in unknown coefficients
of higher-dimensional local operators of the effective theory. For
our considerations, dimension-six operators of the kind $\bar e e 
\bar \nu \nu$ could be dangerous, as they would mimic the graviton
signal. The effect of such an operator suppressed by $M_D^2$
for the $\gamma +\Emiss$
signature at a $1\tev$ $e^+e^-$ collider 
is shown in figs.~\ref{fige1} and \ref{fige2} 
for a particular choice of chirality structure.
Since the overall normalization of this operator is unknown, we
cannot decide on its importance as a source of background. Studies
at different $\sqrt{s}$ and with different beam polarizations would
be essential in discriminating between the signal from graviton
production and the signal from higher-dimensional operators.

\subsection{Hadron Colliders}

We started our phenomenological analysis with the case of $e^+e^-$
colliders because they allow a simpler discussion, as the fundamental
process occurs at a fixed $\sqrt{s}$. We now illustrate
the possibilities at the CERN LHC, which is an approved proton-proton
collider capable of effectively studying the TeV physics region.
The advantage of LHC is the large
centre-of-mass energy available, but potential drawbacks are the
smearing in the effective $\sqrt{s}$ of parton collisions and the
large Standard Model background.

The leading experimental signal of graviton production at the LHC is
$pp\to {\rm jet} +\Emisst$ coming from the subprocess $qg\to qG$ (which
gives the largest contribution),
$q\bar q\to gG$, and $gg\to gG$. The differential
rates for these parton processes have been given in 
eqs.~(\ref{sigqq})--(\ref{siggg}). The main background comes from processes
with a $Z$ boson and one jet in the final state, with the $Z$ decaying
into neutrinos. In hadron colliders we cannot isolate the $Z$-peak
contribution and hence the background, as well as the signal, is given
by two-body processes, in contrast with the case discussed in 
sect.~\ref{sectee}. In figs.~\ref{figp1} we show the 
signal~\cite{grv94}
and background~\cite{pythia}
rates for transverse jet energy larger than 
$E_{T,{\rm jet}}^{\rm min}$, with an
acceptance cut on the jet rapidity $|\eta_{\rm jet}|<3$. 
In this figure we have fixed
$\md =5$ TeV. In fig.~\ref{figp2} we show the $\md$ dependence
for $E_{T,{\rm jet}}^{\rm min}=1$ TeV.
\jfig{figp1}{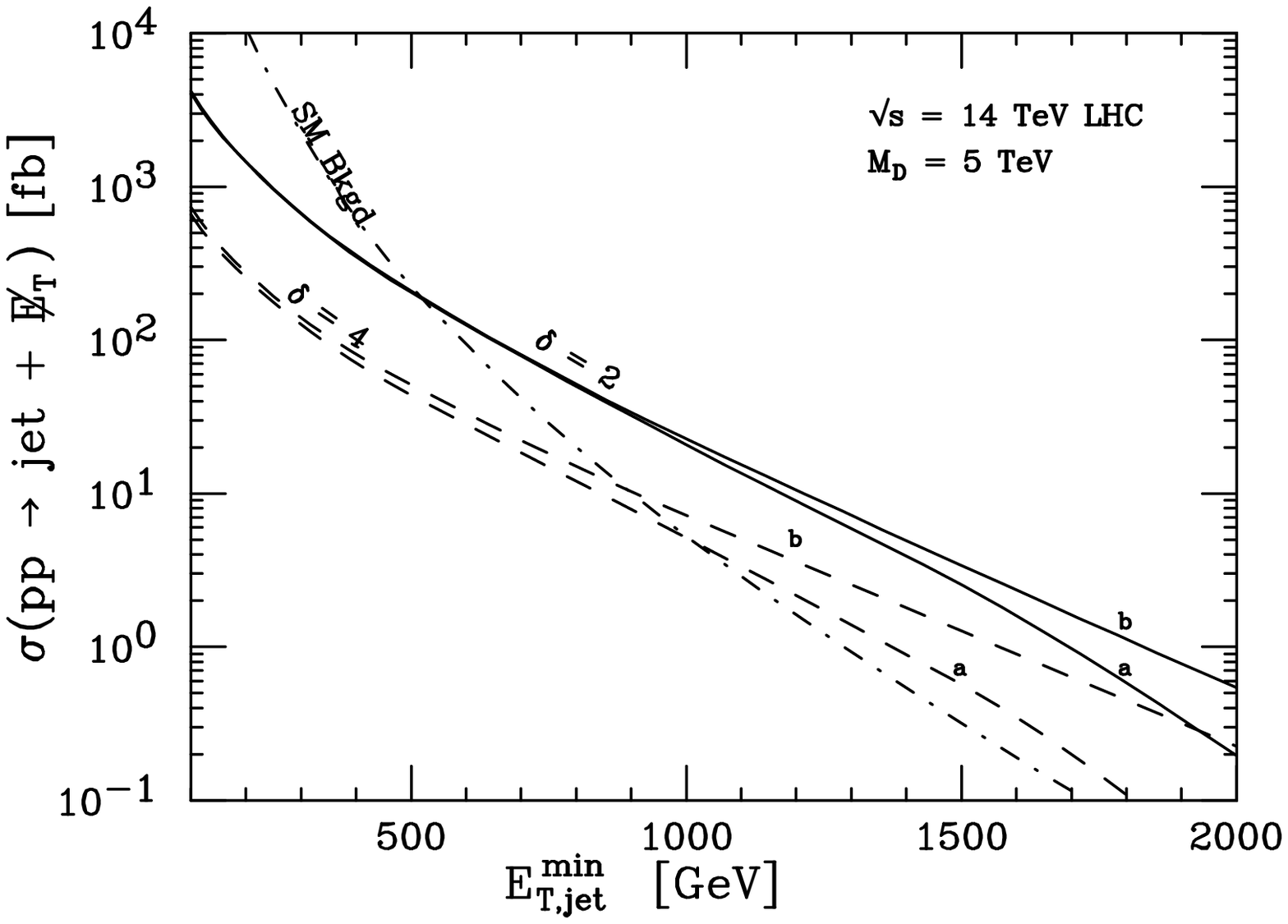}{The total ${\rm jet}+{\rm nothing}$ cross-section
at the LHC integrated for all 
$E_{T,{\rm jet}}> E_{T,{\rm jet}}^{\rm min}$ with the requirement that
$|\eta_{\rm jet}|< 3.0$.  The Standard Model 
background is the dash-dotted line, and the signal
is plotted as solid and dashed lines
for fixed $M_D=5\tev$ with $\delta=2$ and $4$ extra
dimensions.  The {\bf a} ({\bf b}) lines are constructed by integrating
the cross-section over $\hat s < M_D^2$ 
(all $\hat s$).}
\jfig{figp2}{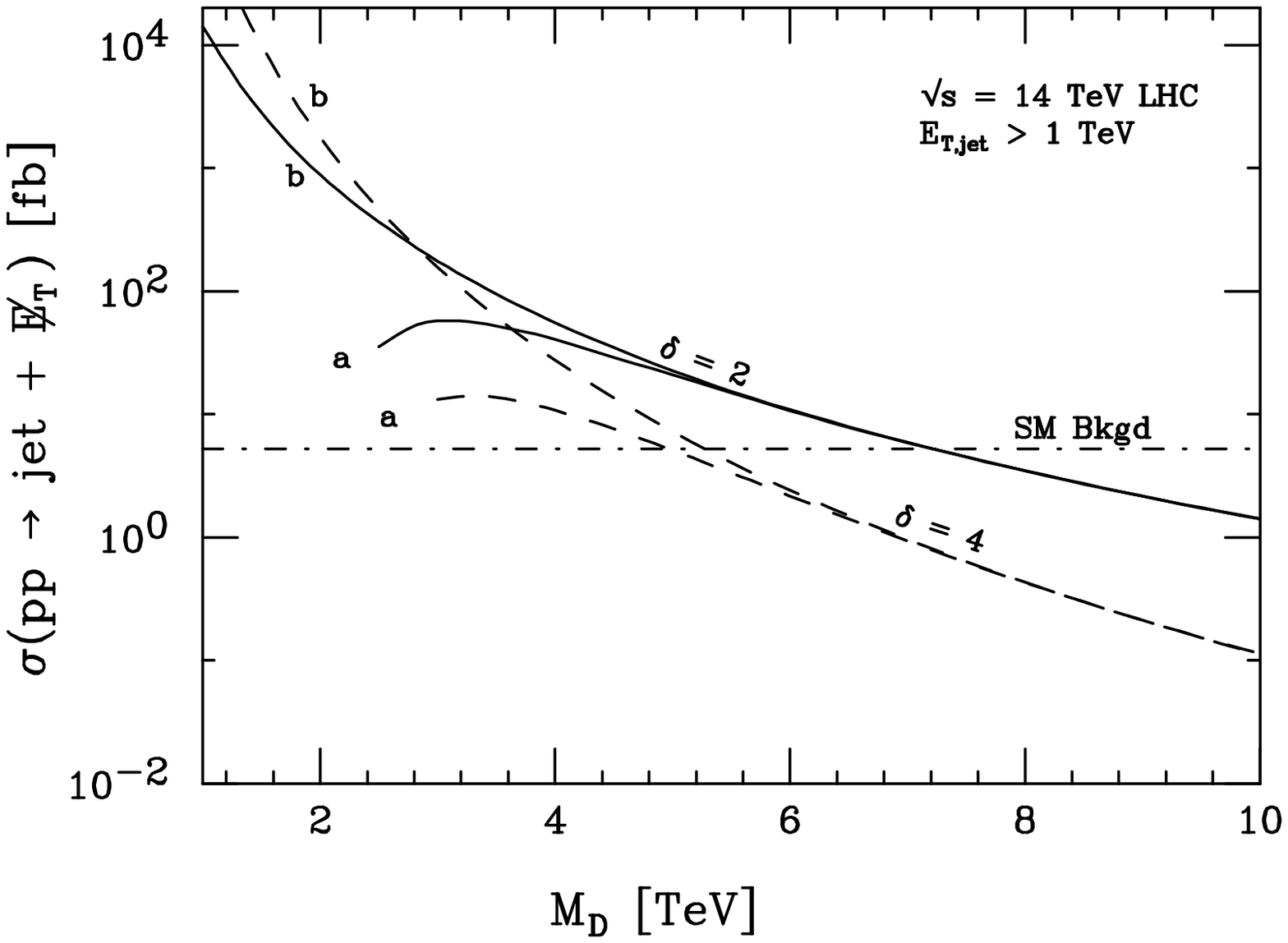}{The total ${\rm jet}+{\rm nothing}$ cross-section
versus $M_D$ at the LHC integrated for all 
$E_{T,{\rm jet}}> 1\tev$ with the requirement that
$|\eta_{\rm jet}|< 3.0$.  The Standard Model background is the 
dash-dotted line, and the signal
is plotted as solid and dashed lines
for $\delta=2$ and $4$ extra
dimensions.  The {\bf a} ({\bf b}) lines are constructed by integrating
the cross-section over $\hat s < M_D^2$ 
(all $\hat s$).}

In hadron colliders the elementary scattering processes occur
at different centre-of-mass energies. Therefore it is not straightforward
to assess the applicability of
the effective-theory approach. 
In order to quantify the ultraviolet sensitivity,
we have plotted in figs.~\ref{figp1} and
\ref{figp2} two curves for each value of $\delta$. 
The first one, denoted by the symbol {\bf a}, is the
result of setting to zero the cross-sections in 
eqs.~(\ref{sigqq})--(\ref{siggg}) whenever the effective centre-of-mass 
energy in the parton collision $\sqrt{\hat s}$ is larger than $\md$,
The second curve, denoted by the symbol {\bf b}, lets the 
cross-sections grow indefinitely with $\sqrt{\hat s}$. In the regions where
the two curves almost coincide, the dominant contribution comes from
momenta smaller than $M_D$ and the effective theory is fully applicable. When
the two curves go apart, the ultraviolet contributions become important,
and our calculation is not under control. 

The existence of a calculable perturbative
region insensitive to the ultraviolet is 
related to the rapid decrease in parton
luminosities with increasing $\sqrt{\hat s}$ which more
than compensates for the increase of the cross-sections in 
eqs.~(\ref{sigqq})--(\ref{siggg}). The larger the value of $\delta$, the
faster the increase in the cross-section, and thus the sooner the
non-perturbative region is reached, as shown in figs.~\ref{figp1} and
\ref{figp2}. 

In order to establish the LHC range of sensitivity,
we consider two options for the integrated luminosity, ${\cal L}=10$~fb$^{-1}$
and 100~fb$^{-1}$. We estimate the systematic error in the background
prediction to be about 10{\%}. This precision
could be reached with a next-to-leading
order calculation or by experimentally calibrating the background to
the measured cross-section for jet + $Z$, with the $Z$ decaying
leptonically $Z\to \ell^+ \ell^-$.

For ${\cal L}=100$~fb$^{-1}$, the systematic error dominates, and 
the sensitivity range is defined by
\beq
\sigma_{\rm Signal} > 5(10\%)\sigma_{\rm Bkgd}
=2.6~{\rm fb}.
\eeq
For ${\cal L}=10$~fb$^{-1}$, the systematic and statistical errors are
comparable. We add them in quadrature and require
\beq
\sigma_{\rm Signal} > \sqrt{2}~\frac{5\sqrt{ 
\sigma_{\rm Bkgd}{\cal L}}}{\cal L}
=3.7~{\rm fb}.
\eeq
The corresponding $\md$ sensitivity
ranges are shown in table~3. These
results are obtained using the curves of type {\bf a}, which give a more
conservative prediction of the signal.
Table~3 also contains an estimate of the lower bound on $\md$
up to which the perturbative calculation presumably can be trusted. 
The criterion used is that the difference between curves {\bf a} and {\bf b}
should be smaller than 50\% of the result in curve {\bf a}. 
It should be realized that by lowering $E_{T,{\rm jet}}^{\rm min}$ the minimum
value of $M_D$ consistent with perturbativity is also lowered.  Therefore,
by adjusting the $E_{T,{\rm jet}}$ 
cut in experimental analyses, one can select different ranges
of $M_D$ to probe perturbatively.
For $\delta \geq
5$, there is no region of $\md$ in which we can simultaneously trust
perturbation theory and obtain a visible signal at LHC.  Nevertheless,
for $\delta <5$ the LHC probes perturbatively the multi-TeV region,
which is prime territory if these ideas ameliorate the fine-tuning
problem between the weak scale and the gravity scale.
In the case of $\delta =2$, the SN1987A bound discussed in sect.~\ref{sectee}
rules out the possibility of an observable signal.

\begin{table}
\centering
\begin{tabular}{cccc}
\hline\hline
   & Max $\md$  & Max $\md$  & Min $\md$ \\
$\delta$    & sensitivity    & sensitivity     & perturbativity \\
   & ${\cal L}=$ 100 fb$^{-1}$    &  ${\cal L}=$ 10 fb$^{-1}$& \\
\hline
2 & 8.5 TeV & 7.9 TeV & 3.8 TeV \\
3 & 6.8~~~~~~  & 6.3~~~~~~  & 4.3~~~~~~  \\
4 & 5.8~~~~~~  & 5.5~~~~~~  & 4.8~~~~~~  \\
5 & 5.0~~~~~~  & 4.6~~~~~~  & 5.4~~~~~~  \\
\hline\hline
\end{tabular}
\caption{{Maximum $\md$ sensitivity which can be reached
by studying the final state ${\rm jet} +\Emisst$ at the LHC
with $\sqrt{s}=14$~TeV and integrated luminosity ${\cal L}=100$~fb$^{-1}$
or 10~fb$^{-1}$.
The bounds have been obtained
by requiring $\sigma_{\rm Signal}>2.6$~fb (for ${\cal L}=100$~fb$^{-1}$)
or 3.7~fb (for ${\cal L}=10$~fb$^{-1}$)
with the acceptance cuts $|\eta_{\rm jet}|<3$ and
$E_{T,{\rm jet}}>1$~TeV. We also give an estimate of the minimum value of $\md$
for which the effective-theory calculation can be trusted.
}}
\end{table}

A different signal for graviton production comes from events with 
a photon and missing energy in the final state, arising from
the subprocess $\bar q  q\to G \gamma$. The Standard Model background
originates mainly from $\bar q  q\to Z \gamma$. Figures~\ref{figp3} and
\ref{figp4} show the results of the
cross-section as a function of the cut on $E_{T,\gamma}$ for a fixed 
$\md$, and as a function of $\md$ for a fixed cut on $E_{T,\gamma}$,
under the requirement $|\eta_\gamma |>2.5$.
\jfig{figp3}{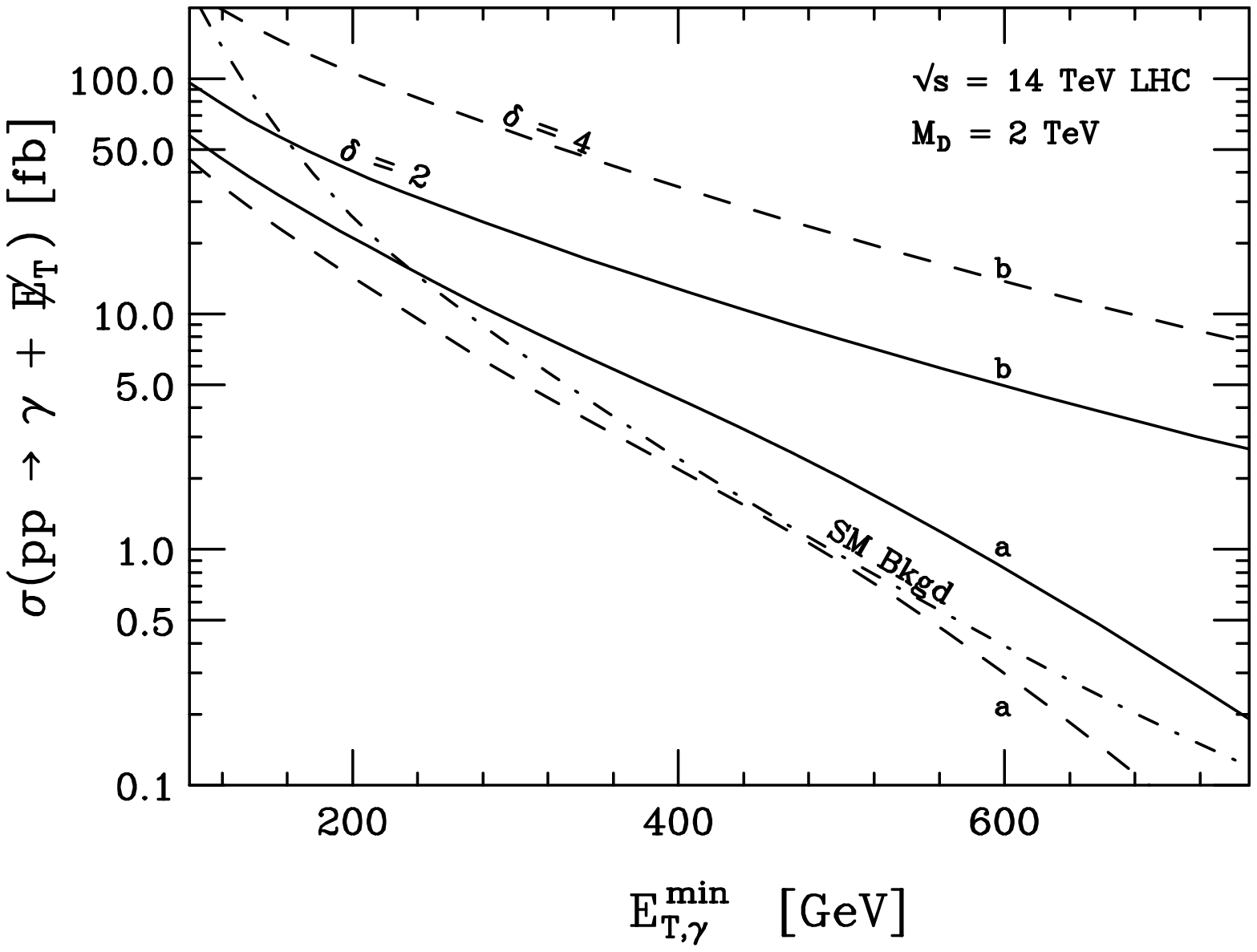}{The total ${\gamma}+{\rm nothing}$ cross-section
at the LHC integrated for all 
$E_{T,\gamma}> E_{T,\gamma}^{\rm min}$ with the requirement that
$|\eta_\gamma| < 2.5$.  The Standard Model background is the 
dash-dotted line, and the signal
is plotted as solid and dashed lines
for fixed $M_D=2\tev$ with $\delta=2$ and $4$ extra
dimensions.  The {\bf a} ({\bf b}) lines are constructed by integrating
the cross-section over $\hat s < M_D^2$ 
(all $\hat s$).}
\jfig{figp4}{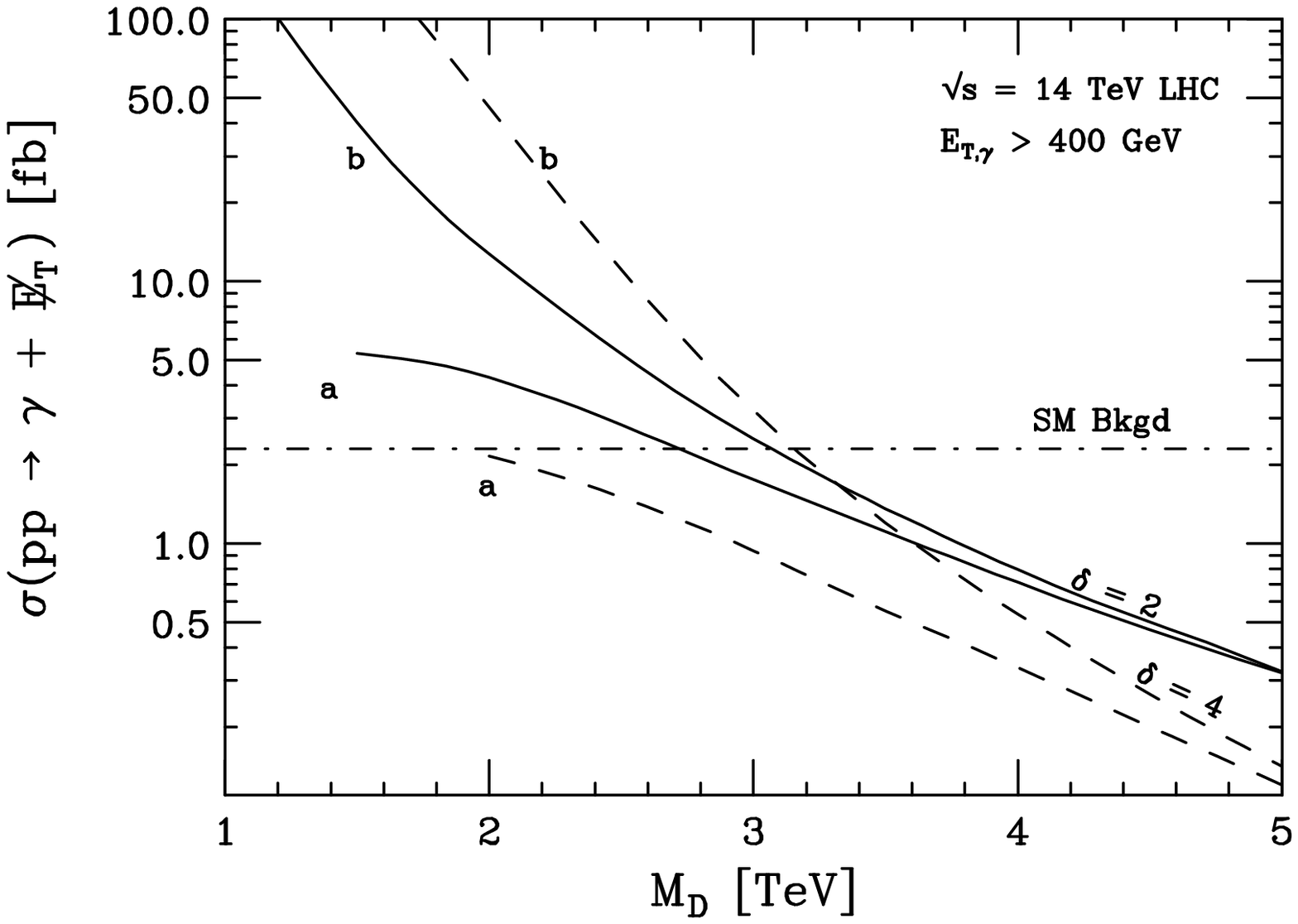}{The total $\gamma +{\rm nothing}$ cross-section
versus $M_D$ at the LHC integrated for all 
$E_{T,\gamma}> 400\gev$ with the requirement that
$|\eta_\gamma| < 2.5$.  The Standard Model background is the 
dash-dotted line, and the signal
is plotted as solid lines
for $\delta=2$ and $4$ extra
dimensions.  The {\bf a} ({\bf b}) lines are constructed by integrating
the cross-section over $\hat s < M_D^2$ 
(all $\hat s$).}

The disadvantage of the photon signal over the jet signal is the much lower
rate, caused by the smallness of the electromagnetic coupling and
the lower luminosity of $\bar q q$ over $q g$ at large
values of $\hat s/s$ in $pp$ colliders. The lower rate requires 
smaller values of $\md$ to achieve a visible signal, and therefore a
much more limited perturbative region. This is illustrated in
figs.~\ref{figp3} and
\ref{figp4}, since the curves {\bf a} and {\bf b} (defined as before) are
significantly separated in most of the interesting region. Therefore,
the sensitivity range of $\md$ obtained from the photon signal is much
smaller than in the jet case. Nevertheless, in case of discovery in
the jet channel, the
photon signal can provide a useful independent test.

As discussed in sect.~\ref{sectee}, other effects inherent to the
full quantum-gravity theory can provide a source of background to the
graviton signal. However, comparing to $e^+e^-$ colliders, these
effects could be less important here. First of all, there are no
dimension-six operators contributing to\footnote{Operators of the
kind $\bar f \gamma_\mu D_\nu f F^{\mu \nu}$ vanish for massless
fermions by the equations of motion.}
$qg\to qZ$, or $\bar q q\to gZ$,
or $gg\to gZ$. Moreover, dimension-six operators of the kind $\bar q q
\bar \nu \nu$ contribute only to processes with three-body final states,
and should be less important than the two-body signal and background
computed above.

We close this section by analysing the capabilities of the Tevatron
in the ${\rm jet}+ \Emisst$ mode.
In fig.~\ref{figpbar1} we plot
the cross-sections~\cite{grv94,pythia} for $p\bar p\to {\rm jet}+\Emisst$ 
at $\sqrt{s}=2\tev$ as a function
of $E^{\rm min}_{T,{\rm jet}}$ with $M_D=1.2\tev$ fixed, and 
$|\eta_{\rm jet}|<3$.  
\jfig{figpbar1}{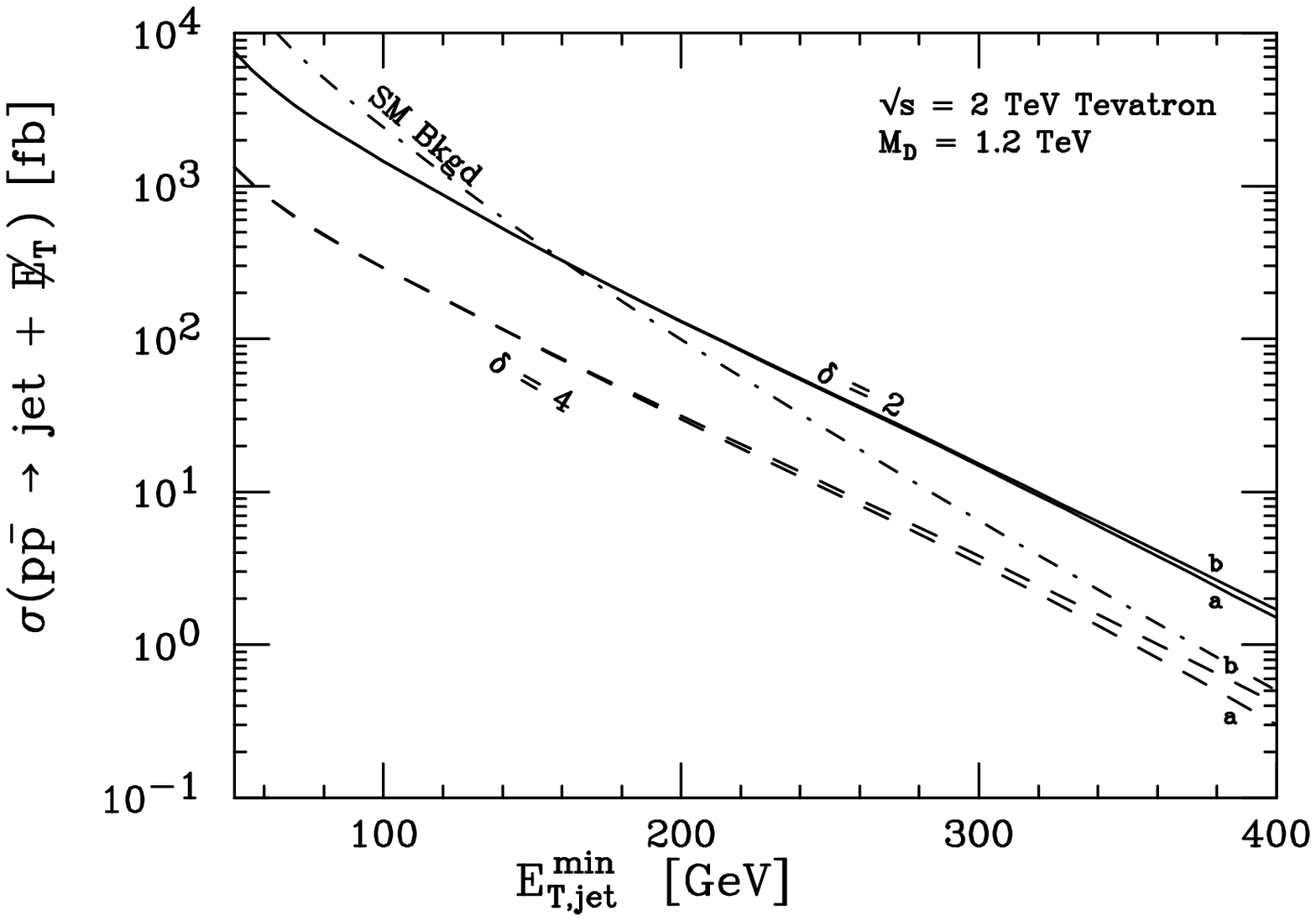}{The total ${\rm jet}+{\rm nothing}$ cross-section
at the Tevatron ($\sqrt{s}=2\tev$) integrated for all 
$E_{T,{\rm jet}}> E_{T,{\rm jet}}^{\rm min}$ with the requirement that
$|\eta_{\rm jet}|< 3.0$.  The Standard Model
background is the dash-dotted line, and the signal
is plotted as solid and dashed lines
for fixed $M_D=5\tev$ with $\delta=2$ and $4$ extra
dimensions.  The {\bf a} ({\bf b}) lines are constructed by integrating
the cross-section over $\hat s < M_D^2$ 
(all $\hat s$).}
Here again, the signal is flatter with increasing~$E_{T,{\rm jet}}^{\rm min}$
than the background, and so it is helpful to go as high in $E_T$ as
is allowed while still having enough events for a statistically significant
signal. In fig.~\ref{figpbar2} we plot the
$M_D$ dependence of the ${\rm jet}+\Emisst$ cross-section 
for $E^{\rm min}_{T,{\rm jet}}=150\gev$.
\jfig{figpbar2}{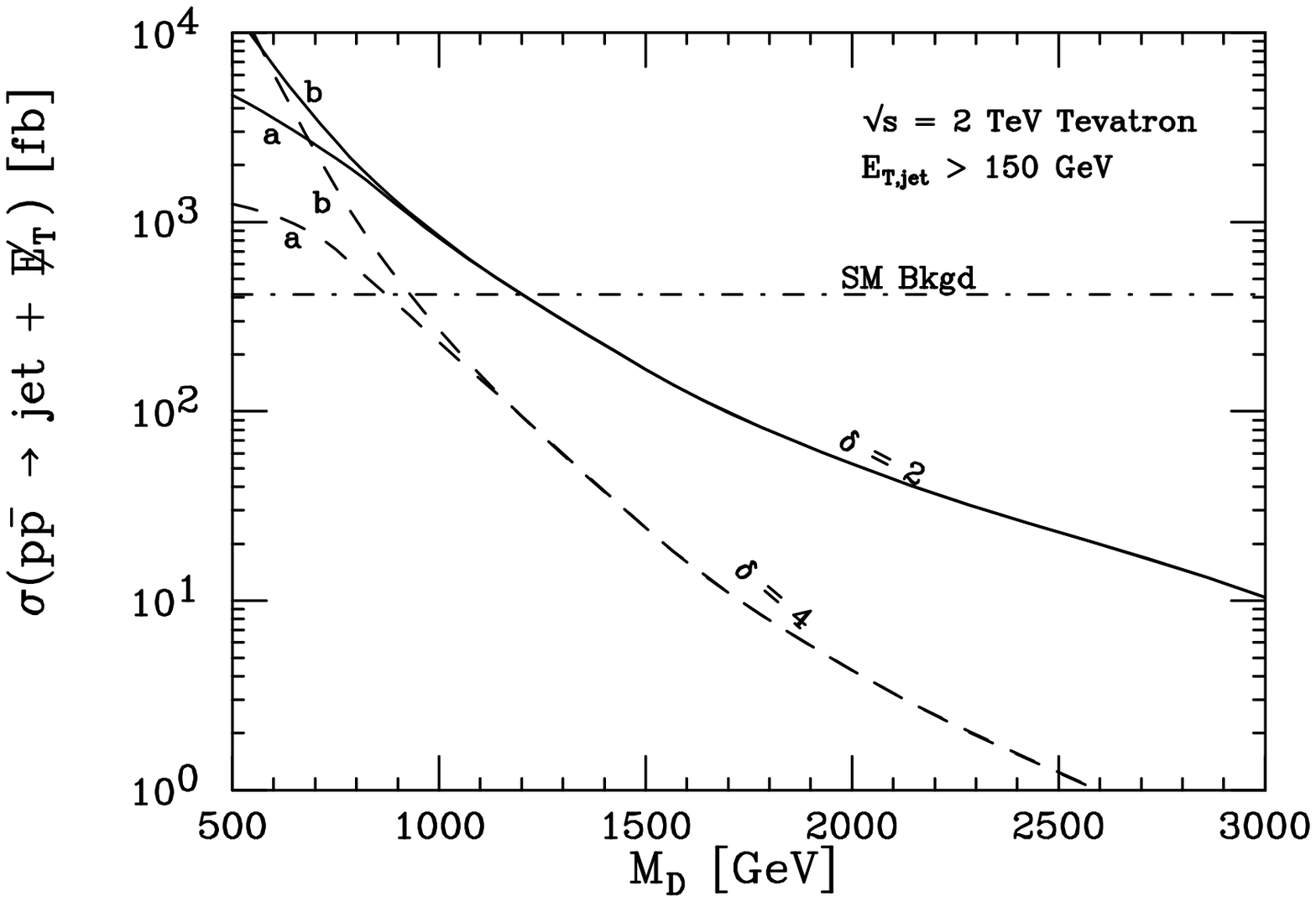}{The total ${\rm jet}+{\rm nothing}$ cross-section
versus $M_D$ at the Tevatron ($\sqrt{s}=2\tev$) integrated for all 
$E_{T,{\rm jet}}> 150\gev$ with the requirement that
$|\eta_{\rm jet}|< 3.0$.  The Standard Model
background is the dash-dotted line, and the signal
is plotted as solid and dashed lines for $\delta=2$ and $4$ extra
dimensions.  The {\bf a} ({\bf b}) lines are constructed by integrating
the cross-section over $\hat s < M_D^2$ 
(all $\hat s$).}

We calculate the $M_D$ sensitivity at the Tevatron for two different
centre-of-mass energies.  For $\sqrt{s}=1.8\tev$,
with $200\xpb^{-1}$ of integrated luminosity, and $E^{\rm min}_{T,{\rm jet}}=
150$ GeV,
the background is approximately
$290\xfb$.  Requiring a signal significance of $S/\sqrt{B}=5$, we find
$\sigma_{\rm sig} > 190\xfb$ for discovery.  Note that this 
is statistics limited,
since we assume that the background is known to better than
about $10\%$.  The resulting sensitivities 
of $M_D$ are listed in table~4.

We carry out the same procedure to calculate sensitivities of $M_D$ for
a $2\tev$ centre-of-mass Tevatron to be available for Run II.
The background with $E^{\rm min}_{T,{\rm jet}}=150$ GeV is
approximately $410\xfb$.
The integrated luminosity for the next run is expected to
be above several $\xfb^{-1}$; however, the added luminosity helps little
when $E^{\rm min}_{T,{\rm jet}}=150$ GeV,
since we are limited by
the systematic uncertainty in knowing the background cross-section.
To take full advantage of the higher luminosity, one can choose a
larger value of $E^{\rm min}_{T,{\rm jet}}$ and gain increasing
sensitivity to $\md$, by enhancing the signal to background ratio.
However, in this case, the minimum value of $\md$ from perturbativity
requirements also increases and therefore, for our illustrative study,
we prefer to keep $E^{\rm min}_{T,{\rm jet}}$ fixed.
If we assume that we will know the background cross-section to 
within $10\%$, either by theory or by inferring it from
$Z\, {\rm jet}\to l^+l^-\, {\rm jet}$,
then confidence of a discovery will require that
the signal be more than $50\%$ of the background.  This systematic
uncertainty is the limiting factor in such searches provided the
integrated luminosity is above
$300\xpb^{-1}$, which is planned to be easily exceeded during Run II.
In table~4 we give the $M_D$ limits according to the preceding
discussion, with $E^{\rm min}_{T,{\rm jet}}=150$ GeV, 
and find sensitivity to $M_D$ above $1\tev$ for
$2\leq \delta \leq 4$.  For $\delta =5$ the signal is not large enough
to be detectable in a purely perturbative regime.  

\begin{table}
\centering
\begin{tabular}{ccc|cc}
\hline\hline
      & Max $\md$  & Min $\md$  & Max $\md$   & Min $\md$  \\
      & sensitivity       & perturbativity          & sensitivity        & perturbativity  \\
$\delta$      & $\sqrt{s}=1.8\tev$  & $\sqrt{s}=1.8\tev$     & $\sqrt{s}=2\tev$ & $\sqrt{s}=2\tev$ \\
      & ${\cal L}=200\xpb^{-1}$ &   & ${\cal L}>300\xpb^{-1}$ &  \\
\hline
2 & 1100 GeV & 600 GeV & 1400 GeV & 650 GeV \\
3 &  950~~~~~~ & 700~~~~~~ & 1150~~~~~~ & 750~~~~~~ \\
4 & 850~~~~~~ & 800~~~~~~ & 1000~~~~~~ & 850~~~~~~ \\
5 & 700~~~~~~ & 900~~~~~~ & 900~~~~~~ & 950~~~~~~ \\
\hline\hline
\end{tabular}
\caption{{Maximum $\md$ sensitivity which can be reached
by studying the final state ${\rm jet} +\Emisst$ at the Tevatron
with $\sqrt{s}=1.8$~TeV and integrated luminosity 
${\cal L}=200\xpb^{-1}$, or $\sqrt{s}=2\tev$ and integrated
luminosity ${\cal L}>300\xpb^{-1}$. 
The bounds have been obtained
by requiring $\sigma_{\rm Signal}>190$~fb (for $\sqrt{s}=1.8\tev$)
or 205~fb (for $\sqrt{s}=2\tev$)
with the acceptance cuts $|\eta_{\rm jet}|<3$ and
$E_{T,{\rm jet}}>150$~GeV. We also give an 
estimate of the minimum value of $\md$
for which the effective-theory calculation can be trusted.}}
\end{table}

\section{The Operator ${\cal T}$ and Collider Experiments}
\label{secop}

As we have shown in sect.~\ref{secvir}, the operator ${\cal T}$ defined in
eq.~(\ref{top}) is generated by a one-graviton exchange at tree level.
Its coefficient in the interaction Lagrangian 
\beq
{\cal L}=\frac{4\pi}{\Lambda_T^4}~{\cal T}
\eeq
is an ultraviolet-dependent
unknown parameter, although the non-analytic energy dependence 
can be computed
with the low-energy
effective theory (see sect.~\ref{secvir}). 
We will proceed on our phenomenological analysis assuming that
$\Lambda_T$ is an energy-independent parameter of order $\md$.

Since the coefficient of the dimension-8
operator ${\cal T}$ is sensitive to
ultraviolet physics, it is certainly model dependent. However, we want to
argue here that the operator ${\cal T}$ has a special meaning. Although
graviton loops (or other quantum-gravity effects)
can produce any generic operator, even those of dimensions
less than 8, ${\cal T}$ is generated at tree level. If momenta integrals
are cutoff at a scale equal to
or less than $\md$, the loop suppression may appear
in the coefficients of the operators in the effective theory.
Moreover, the operator ${\cal T}$ gives a pure $d$-wave (spin two)
contribution to certain scattering processes, a signature of the graviton
existence. The same operator simultaneously gives contributions to several
scattering processes, offering the possibility of testing gravitational
universality. Finally, for certain processes, it is possible to show that no
other operator of lower dimension can compete with ${\cal T}$.
For all these reasons, we believe that it is phenomenologically 
interesting to focus briefly on experimental tests of ${\cal T}$.

\subsection{$e^+e^-$ and Muon Colliders}

The operator ${\cal T}$ gives new contributions to Bhabha scattering
in $s$, $t$, and $u$ channels. It gives pure $d$-wave contributions
to $e^+e^-\to \mu^+\mu^-$, to vector boson production
($e^+e^-\to \gamma \gamma , ZZ,W^+W^-$),
 and to processes with two-jet final states
($e^+e^-\to \bar q q$, $e^+e^-\to gg$), three jets ($e^+e^-\to \bar q q
g$, $e^+e^-\to ggg$), and four jets ($e^+e^-\to gggg$). Another
interesting process is Higgs pair production $e^+e^-\to HH$,
which has a cross-section independent of the electron Yukawa coupling.
In absence of other operators, the deviations
{}from the Standard Model predictions for all these processes are determined
in terms of a single parameter $\Lambda_T$. In practice, it will be
a highly non-trivial experimental task to disentangle many new contributions.

The most promising process is
$e^+e^-\to \gamma \gamma$. Indeed, there are no operators
of dimension less than 8 which can contribute
at tree level to
this process\footnote{The dimension-8 operator $\bar f \gamma^\mu
D_{\{ \nu} D_{\lambda \} }f D_\nu F_{\mu \lambda}$
is independent of
${\cal T}$ and contributes to $e^+e^-\to \gamma \gamma$. On the other hand,
using the Bianchi identity, it can be shown
that the operator $\bar f \gamma^\mu f F^{\nu \lambda}
D_\lambda F_{\mu \nu}$ vanishes under the equations of motion.}.
Therefore, observables associated with the two-photon final state may be
good probes of $\Lambda_T$.  Angular correlations of the photons with
the electron have been used effectively to search for excited electron states
contributing in the $t$-channel to $e^+e^-\to \gamma\gamma$~\cite{yyexpts}.
A large composite electron mass in these studies is analogous to $\Lambda_T$.
Therefore, the techniques employed to detect effects of compositeness can be
directly applied to study effects of the local operator ${\cal T}$
on $e^+e^-\to \gamma\gamma$.  

A complete experimental study of this type
would require scrutiny of the differential angular distribution
of the photons.  Binned counts of the photons produced at different angles
are then compared statistically
with the Standard Model expected distribution to detect
anomalous behaviour.  We do not present such a study here, but rather
simplify the analysis to first plotting the integrated $E_T$ spectrum of the
photons, then choosing a value of $E_{T,\gamma}^{\rm min}$ and comparing
total counts expected in the Standard Model to those predicted for a given 
value of $\Lambda_T$. 

In fig.~\ref{fige3} we plot the $e^+e^-\to \gamma\gamma$ cross-section
for different values of $\Lambda_T$ as a function of $E^{\rm min}_{T,\gamma}$
for a $1\tev$ $e^+e^-$ collider.
\jfig{fige3}{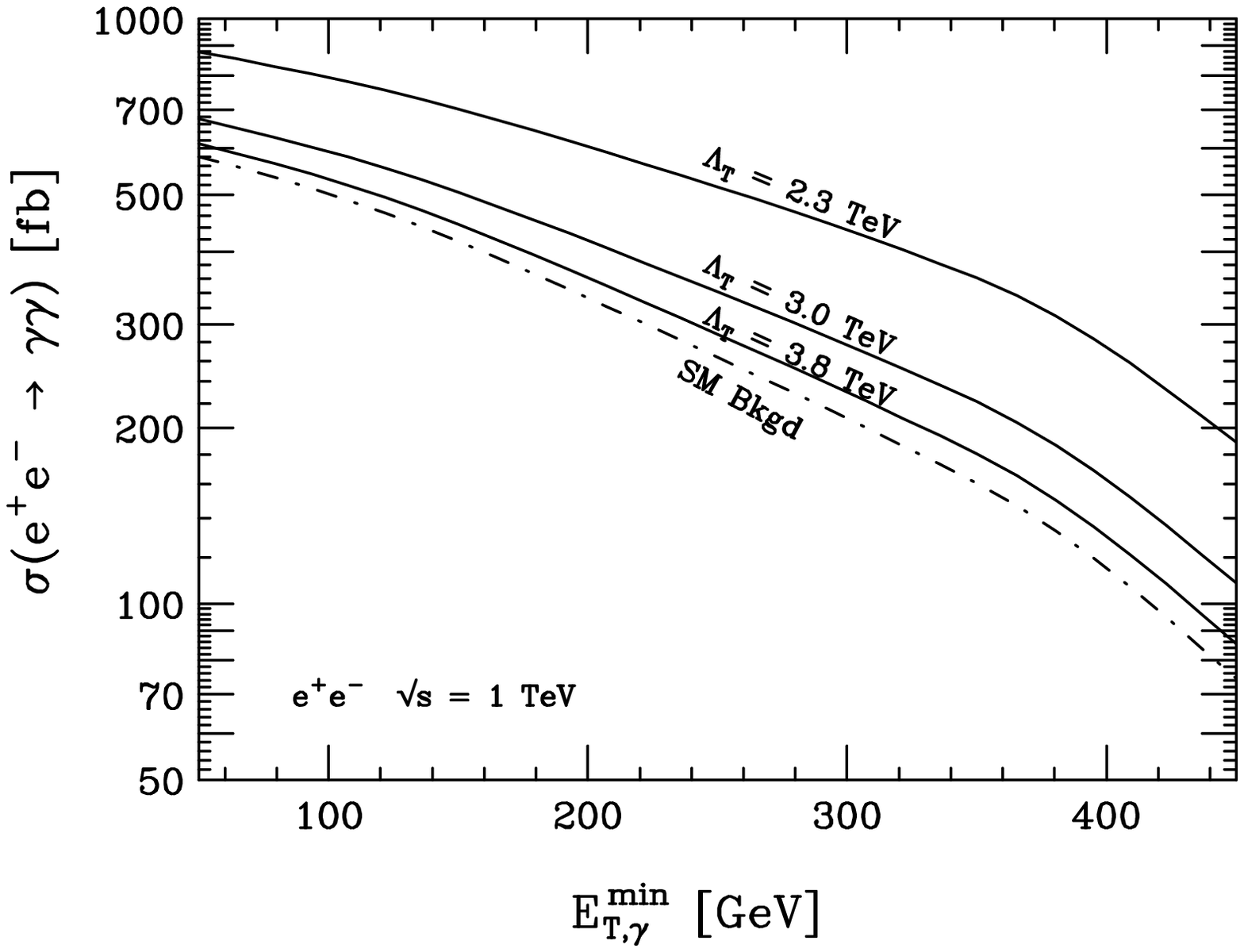}{Total cross-section for $e^-e^+\to \gamma\gamma$
versus $E_{T,\gamma}^{\rm min}$ at a $1\tev$ centre-of-mass energy collider.  
The
dashed line represents the Standard Model 
background, and the solid lines represent
the total cross-section for various values of $\Lambda_T$.}
As expected, the higher in $E_{T,\gamma}$ we search, the more the signal
wins over the background.  If we require, for example, that 
$E_{T,\gamma}> 300\gev$, and also require for discovery that the total
signal must be more than $10\%$ of the background, we find
sensitivity to $\Lambda_T$ up to $3.8\tev$.  This limit
on $\Lambda_T$ is independent of the number of extra dimensions; however,
it should be noted that
the relationship between $\Lambda_T$ and $M_D$ likely depends on $\delta$
in an incalculable way.

Similar searches can also be performed at LEP. An interesting peculiarity
is the study of virtual-graviton contributions to observables at the
$Z$ peak. Interference effects come only from imaginary parts and
therefore from gravitons with masses equal to $\sqrt{s}$.
These contributions can be computed reliably using
the effective theory alone and can affect 
angular distributions of two-fermion final states. 
We estimate, however, that the searches for
graviton production at LEP2 described in sect.~6.1 have better
sensitivities to $\md$.

\subsection{Hadron Colliders}

Similar to the case of $e^+e^-$ colliders, the two-photon final state
provides an interesting signal of the ${\cal T}$ operator at hadron 
colliders.
The $\gamma\gamma$ final state has another advantage
in the hadron collider environment:  the 
measurable invariant mass of the two photons allows investigation
of a possible signal
in different energy domains (different
$\hat s$ regions), thereby enabling one to study the scaling
behaviour of observables sensitive to virtual gravitons.  

For the present, we show the effects of the operator ${\cal T}$ on
$pp\to \gamma\gamma$ collisions by plotting in fig.~\ref{figp5}
the integrated cross-section 
versus $M^{\rm min}_{\gamma\gamma}$ for several values  of $\Lambda_T$. 
\jfig{figp5}{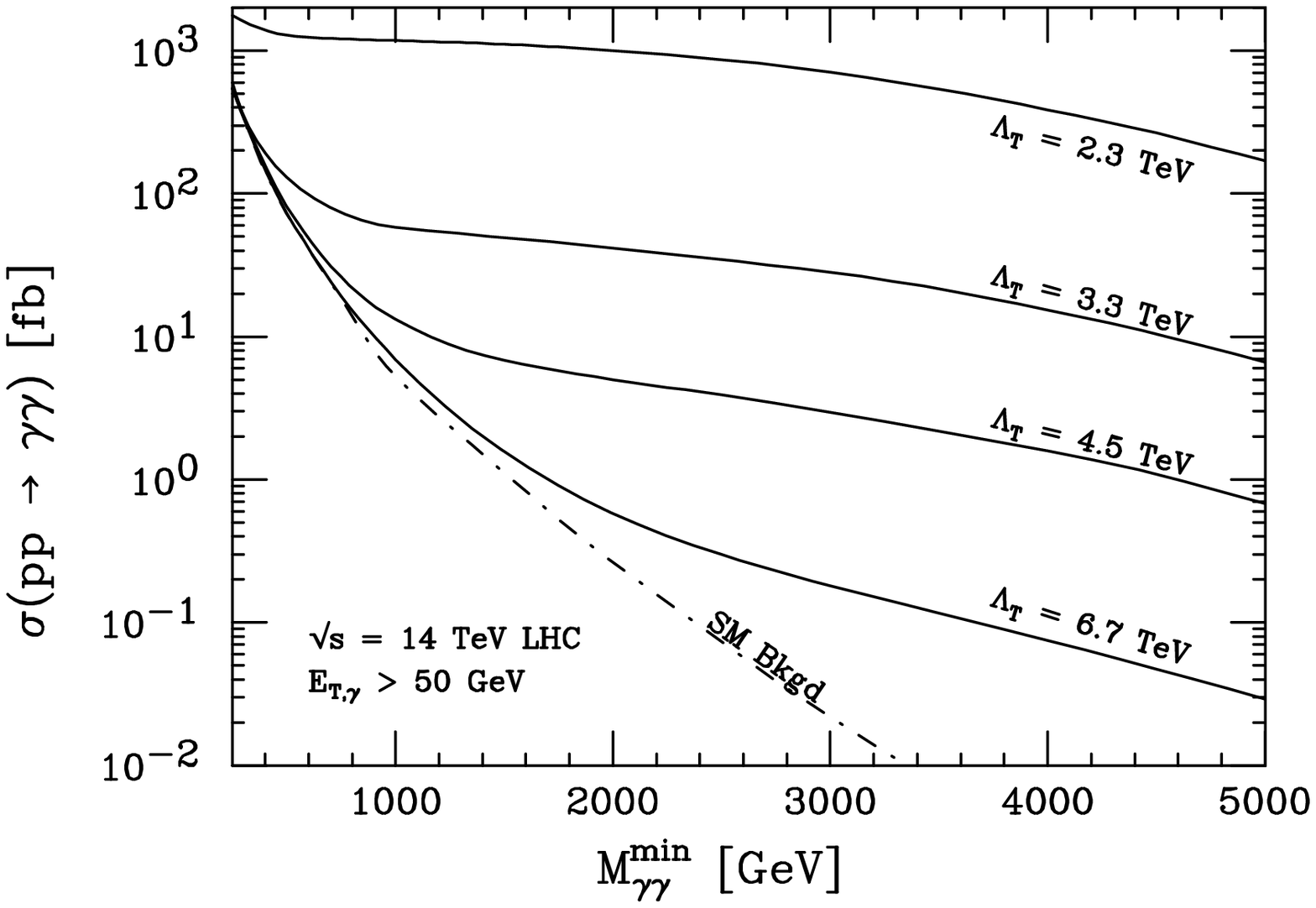}{The total cross-section for $pp\to \gamma\gamma$
integrated for $\sqrt{\hat s} > M_{\gamma\gamma}^{\rm min}$ with
the requirement that $E_{T,\gamma}>50\gev$ and $|\eta_\gamma |< 2.5$ 
for each photon.  The dashed line is the Standard Model
background and the solid
lines are the total cross-sections for various values of $\Lambda_T$.}
This is done for the $14\tev$ LHC, 
requiring $|\eta_\gamma| < 2.5$ and $E_{T,\gamma}>50\gev$ for both photons.
We find that the signal to background 
ratio is
enhanced by going to higher $M^{\rm min}_{\gamma\gamma}$.  
However, keeping typical
LHC luminosities in mind, we employ the cut 
$M^{\rm min}_{\gamma\gamma}>2\tev$ to our
signal and background to make sure enough events are detectable.
The Standard Model background is approximately $0.25\xfb$ for this
choice of $M_{\gamma\gamma}^{\rm min}$.
With $10\xfb^{-1}$ ($100\xfb^{-1}$) we require 10 (50) signal events as
discovery criteria, leading to sensitivity of $\Lambda_T$ up
to $5.8\tev$ ($7.1\tev$).  Again, this limit is independent of 
the number of extra dimensions,
but the relationship between $\Lambda_T$ and $M_D$ is likely not.

\section{Conclusions}

In this paper we have studied collider tests of the idea that gravity
propagates in extra dimensions with very large radii~\cite{dim}. Production
of \kk excitations of the graviton can be predicted in a fairly
model-independent way, using an effective Lagrangian approach.
Here we have discussed the formalism to derive such an effective theory.

The basic 
collider signal of graviton production is missing energy, accompanied
by a gauge boson or a hadronic jet. The graviton 
signal can be distinguished
from other new-physics sources of similar events by studying different
final-state topologies and kinematic distributions. Here we have 
discussed the discovery reach of LEP2, the Tevatron,
the LHC, and future $e^+e^-$ and
muon colliders.   It is worth emphasizing that the effective
Lagrangian approach is valid only in a restricted energy window
for which a signal is discernible. Therefore,
the existence of various future high-energy colliders can be vital for
this search, since the relevant energy window is different in 
different experiments. 
For instance, LHC could discover new physics phenomena specific
to quantum gravity, and an $e^+e^-$ collider could disentangle the
many possible contributions to related signals, 
and measure the energy and angular dependent
scalings of the graviton-induced cross-sections.

We have also studied simple scattering processes which receive contributions
from virtual-graviton exchange at tree level. Deviations from Standard
Model predictions could be an indication of the presence of \kk gravitons,
although this signal is more model dependent.

The use of the effective theory is well justified in the infrared, but we
cannot predict reliably 
the ultraviolet cutoff up to which we can extrapolate our
results. At high energies,
other unknown phenomena can appear and mimic our signal. Therefore, in the
absence of anomalous events, our analysis can be directly applied to
obtain lower bounds on the quantum-gravity scale. However, in case of
discovery, careful experimental tests and studies of the correlations among
new phenomena would need to be performed to obtain an understanding of
the underlying theory.
Nevertheless, it is very interesting that colliders originally
planned to study the electroweak scale could give us direct information
on the structure of quantum gravity.

\bigskip

\noindent
{\it Acknowledgements: }
We have greatly benefitted from discussions with S.~Dimopoulos, M.~Mangano,
M.~Moretti, M.~Porrati, A.~Signer, R.~Sundrum, Z.~Was, and A.~Zaffaroni.

\section*{Appendix}
\bigskip

The $F_i(x,y)$ functions are
\bea
F_1(x,y)&=&
\frac{1}{x(y-1-x)}\left[ -4x(1+x)(1+2x+2x^2)+ \right.
\nonumber \\&& \left. y(1+6x+18x^2+16x^3)-6y^2x(1+2x)+y^3(1+4x)\right], \\
F_2(x,y)&=& -(y-1-x)~F_1\left( \frac{x}{y-1-x}, \frac{y}{y-1-x}\right) =
\nonumber \\ &&
\frac{1}{x(y-1-x)}\left[ -4x(1+x^2) +y(1+x)(1+8x+x^2)
\right. \nonumber \\ && \left.
-3y^2(1+4x+x^2)
+4y^3(1+x) -2y^4 \right], \\
F_3(x,y)=&&\frac{1}{x(y-1-x)}
\left[ 1+2x+3x^2+2x^3+x^4
\right. \nonumber \\ && \left.
-2y(1+x^3)+3y^2(1+x^2)-2y^3(1+x)+y^4 \right] .
\eea
The function $F_1(x,y)$ determines the cross-section for
$f \bar f \to \gamma G$.
Because of QED invariance under charge conjugation, 
$F_1(x,y)$ is
invariant under exchange of the Mandelstam variables $t$ and $u$.
This is reflected in the property $F_1(x,y)=F_1(y-1-x,y)$. The same
property holds also for the function $F_3(x,y)$, relevant to
the QCD process.

The $G_i(x)$ functions are,
\bea
G_1(x)&=& \left[ \frac{1+2x+2x^2}{-x(1+x)}\right]^{1/2}, \\
G_2(x)&=& \left[ \frac{-x}{16}(1+x)(1+2x+2x^2)\right]^{1/2}, \\
G_3(x)&=& 1+4x+6x^2+4x^3+2x^4, \\
G_4(x)&=& 1+10x+42x^2+64x^3+32x^4, \\
G_5(x)&=& 1+6x+12x^2+8x^3, \\
G_6(x)&=& 1+6x+6x^2, \\
G_7(x)&=& 9x^{-1}+22+24x+11x^2+x^3 , \\
G_8(x)&=& 4+9x+6x^2+x^3, \\
G_9(x)&=& 9+18x+15x^2+5x^3, \\
G_{10}(x)&=& 1+12x+15x^2+5x^3, \\
G_{11}(x)&=& 40+114x+126x^2+60x^3+9x^4.
\eea
The functions $G_1,G_2,G_3,G_4,G_6$ are also invariant under exchange
of $t$ and $u$ and therefore satisfy the property $G_i(-1-x)=G_i(x)$.


\end{document}